\documentclass[usenatbib,useams]{mn2e}

\pdfoutput=1

\usepackage{times}
\usepackage[fleqn]{amsmath}  % needed for boldsymbol
\usepackage{amssymb}
\usepackage{xfrac} % needed for sfrac
\usepackage{aas_macros}
\usepackage{graphicx}

\usepackage{color}
\newcommand{\wcen}{$\omega$~Centauri}
\newcommand{\kms}{km\,s$^{-1}$}
\newcommand{\masyr}{mas\,yr$^{-1}$}

\def\vx{v_{\mathrm{x}}}
\def\vy{v_{\mathrm{y}}}
\def\vz{v_{\mathrm{z}}}
\def\vxp{v_{\mathrm{x'}}}
\def\vyp{v_{\mathrm{y'}}}
\def\vzp{v_{\mathrm{z'}}}
\def\vR{v_{\mathrm{R}}}
\def\vphi{v_{\mathrm{\phi}}}
\def\vlos{v_{\mathrm{los}}}

\def\vzbar{\overline{\vz}}
\def\vxpbar{\overline{\vxp}}
\def\vypbar{\overline{\vyp}}
\def\vzpbar{\overline{\vzp}}
\def\vRbar{\overline{\vR}}
\def\vphibar{\overline{\vphi}}

\def\vxsqbar{\overline{\vx^2}}

\def\vzsqbar{\overline{\vz^2}}
\def\vxpsqbar{\overline{\vxp^2}}
\def\vypsqbar{\overline{\vyp^2}}
\def\vzpsqbar{\overline{\vzp^2}}
\def\vRsqbar{\overline{\vR^2}}
\def\vphisqbar{\overline{\vphi^2}}

\def\vxpvypbar{\overline{\vxp\vyp}}
\def\vxpvzpbar{\overline{\vxp\vzp}}
\def\vypvzpbar{\overline{\vyp\vzp}}
\def\vRvzbar{\overline{\vR\vz}}

\def\sumk{\sum_{k=1}^{N}}
\def\sumj{\sum_{j=1}^{M}}

\def\nuk{\nu_{k}}
\def\nuok{\nu_{0k}}

\def\rhooj{\rho_{0j}}

\def\hju{\mathcal{H}_{j}(u)}

\def\sksq{\sigma_k^2}
\def\sjsq{\sigma_j^2}

\def\qju{1 - (1 - q_j^2) u^2}

\def\kappak{\kappa_k}
\def\kappaksq{\kappa_k^2}

\def\df{\; \; \mathrm{d}}

\def\aa{\mathcal{A}}
\def\bb{\mathcal{B}}
\def\cc{\mathcal{C}}
\def\dd{\mathcal{D}}
\def\ee{\mathcal{E}}
\def\ff{\mathcal{F}}
\def\gg{\mathcal{G}}
\def\ii{\mathcal{I}}
\def\kk{\mathcal{K}}

\def\fabk{\ff_{\alpha \beta, k}}
\def\iabk{\ii_{\alpha \beta, k}}

\def\ab{\left( \aa + \bb \right)}

\def\rte{\ee^{\frac{1}{2}}}

\def\rteqju{\sqrt{\ee \left[ \qju \right]}}

\def\fpigu{4 \pi^{\frac{3}{2}} G \int_0^1 \sumk \sumj \;}
\def\nqpui{\nuok q_j \rhooj u^2 \iabk}
\def\vvdenom{\frac{1}{\left( \cusq \right) \rteqju}}
\def\expabxy{\exp \left[ - \aa \left( x'^2 + \frac{\ab}{\ee} y'^2 \right)
    \right]}

\def\nuvrsqk{\left[ \nu \vRsqbar \right]_k}
\def\nuvphisqk{\left[ \nu \vphisqbar \right]_k}
\def\nuvzsqk{\left[ \nu \vzsqbar \right]_k}

\def\nuvphik{\left[ \nu \vphibar \right]_k}

\def\intinf{\int_{-\infty}^{\infty}}

\def\cusq{1 - \cc u^2}

\def\Msun{\rm M_\odot}
\def\Lsun{\rm L_\odot}

\def\qbar{\overline{q}}

\def\bxi{\mathbf{x'}_i}
\def\vi{\mathbf{v}_i}
\def\vxi{v_{x',i}}
\def\vyi{v_{y',i}}
\def\vzi{v_{z',i}}

\def\svxi{\sigma_{v_{x'},i}}
\def\svyi{\sigma_{v_{y'},i}}
\def\svzi{\sigma_{v_{z'},i}}

\def\vxm{\overline{v_{\mathrm{x'}}}}
\def\vym{\overline{v_{\mathrm{y'}}}}
\def\vzm{\overline{v_{\mathrm{z'}}}}

\def\vsqxm{\overline{v^2_{\mathrm{x'}}}}
\def\vsqym{\overline{v^2_{\mathrm{y'}}}}
\def\vsqzm{\overline{v^2_{\mathrm{z'}}}}

\def\vsqxym{\overline{v^2_{\mathrm{x'y'}}}}
\def\vsqxzm{\overline{v^2_{\mathrm{x'z'}}}}
\def\vsqyzm{\overline{v^2_{\mathrm{y'z'}}}}

\def\bTheta{\boldsymbol{\Theta}}
\def\bmu{\boldsymbol{\mu}}
\def\bC{\mathbf{C}}

\def\Si{\mathbf{S}_i}

\def\bThcl{\bTheta^{\mathrm{cl}}_j}
\def\bThbg{\bTheta^{\mathrm{bg}}_j}

\def\bCcl{\bC^{\mathrm{cl}}_{ij}}
\def\bCbg{\bC^{\mathrm{bg}}_{ij}}

\def\bmucl{\bmu^{\mathrm{cl}}_{ij}}
\def\bmubg{\bmu^{\mathrm{bg}}_{ij}}

\def\Lj{\mathcal{L}_j}
\def\Lij{\mathcal{L}_{ij}}
\def\Lijcl{\mathcal{L}_{ij}^{\mathrm{cl}}}
\def\Lijbg{\mathcal{L}_{ij}^{\mathrm{bg}}}

\def\lj{\ell_j}

\def\pmi{m_i}

\usepackage[pdftex,bookmarks=true]{hyperref}

\hypersetup{
    pdftitle={Discrete dynamical models of \wcen{}},
    pdfauthor={Laura L. Watkins, Glenn van de Ven, Mark den Brok, Remco C. E. van den Bosch}
}

\def\equationautorefname~#1\null{%
  equation~(#1)\null
}

\title[Discrete dynamical models of \wcen{}]{Discrete dynamical models of \wcen{}}

\author[L.L. Watkins et al.]{Laura~L.~Watkins$^{1}$\thanks{watkins@mpia.de},
    Glenn~van~de~Ven$^1$, Mark~den~Brok$^2$, Remco~C.~E.~van~den~Bosch$^1$ \\
    $^1$Max Planck Institute for Astronomy, K\"{o}nigstuhl 17, 69117 Heidelberg, Germany\\
    $^2$Department of Physics and Astronomy, University of Utah, Salt Lake City, UT 84112, USA}

\date{Accepted ???; Received ???; Submitted ???}

\pagerange{\pageref{firstpage}--\pageref{lastpage}} \pubyear{2013}

\begin{document}

\label{firstpage}

\maketitle

\begin{abstract}
    We present a new framework for modelling discrete kinematic data.  Current techniques typically involve binning.  Our approach works directly with the discrete data and uses maximum-likelihood methods to assess the probability of the dataset given model predictions.  We avoid making hard cuts on the datasets by allowing for a contaminating population in our models.  We apply our models to discrete proper-motion and line-of-sight-velocity data of Galactic globular cluster \wcen{} and find a mildly radial velocity anisotropy $\beta = 0.10 \pm 0.02$, an inclination angle $i = 50^\circ \pm 1^\circ$, a V-band mass-to-light ratio $\Upsilon = 2.71 \pm 0.05 \; \Msun / \Lsun$ and a distance $d = 4.59 \pm 0.08$~kpc.  All parameters are in agreement with previous studies, demonstrating the feasibility of our methods.  We find that the models return lower distances and higher mass-to-light ratios than expected when we include proper motion stars with high errors or for which there is some blending.  We believe this not a fault of our models but is instead due to underestimates or missing systematic uncertainties in the provided errors.
\end{abstract}


% -----------------------------------------------------------------------------

\section{Introduction}
\label{sect:introduction}

The structure and formation history of a system of stars is encoded in its kinematics.  Stars trace the underlying potential in which they orbit; by studying their dynamics, we can determine how much mass is present and where that mass is located.

Globular clusters and dwarf spheroidal galaxies are among the lowest-mass stellar systems in the Universe.  They are spherical or mildly flattened and span a similar range of absolute magnitudes, although dwarf spheroidals are larger than globular clusters of comparable brightness.  Dwarf spheroidals show evidence of multiple stellar populations with complex star formation histories \citep[e.g.][]{deboer2012}.  Globular clusters were once held up as prototypical single stellar populations, however, recent studies have shown that they too appear to host multiple stellar populations \citep[see e.g.][]{gratton2004, piotto2009, piotto2012}.  Kinematical studies reveal that globular cluster dynamics can be explained by accounting for the mass contained in stars, while dwarf spheroidal dynamics require significant amounts of dark matter to reconcile their mass budget.  We do not yet understand how objects so apparently similar at first glance can have such different underlying kinematics; nor it is clear how these objects form and evolve.

For example, one formation mechanism for globular clusters supposes that they are the stripped nuclei of dwarf galaxies \citep{freeman1993, bekki2003}.  Dark matter has yet to be detected in any globular cluster, but theory predicts that if they really are stripped nucleated dwarf ellipticals then there should be small amounts of dark matter left for us to find.  We turn to the internal dynamics of the clusters.  By studying the kinematics of the stars, we are able to constrain the total mass and, from photometric studies, we can determine how much mass is present in their baryonic components.  In principle, provided the uncertainties on our estimates are small, the difference between the baryonic mass and the total mass can then tell us how much dark matter is present in these systems.  Detailed dynamical modelling of the innermost parts of these objects can also detect intermediate mass black holes at their centres, if indeed they are present there.  However, once again, detection is sensitive to the uncertainties in the results, and at present, no intermediate mass black hole has been robustly detected in any globular cluster.

Similar analyses can also be applied to dwarf spheroidal galaxies.  We know that these objects are dark-matter dominated, with mass-to-light ratios $M/L_{\mathrm{V}} \sim 10^{1-3}$ \citep[e.g.][]{mateo1998, simon2011}, but the distribution of the dark matter is still an open question.  $\Lambda$CDM simulations predict that dwarf spheroidals should have cuspy dark matter profiles \citep{navarro1996}; observations tend to favour cored profiles but are unable to completely rule out cusps \citep[e.g.][]{amorisco2012a, jardel2012, breddels2012}.  Recently, \citet{pontzen2012} showed that dark matter cusps can become cores as a result of supernova feedback following centrally concentrated bursts of intense star formation.  However, this requires efficient star formation, while the deficit of Galactic satellites compared against $\Lambda$CDM predictions (the ``missing satellite problem") implies that star formation is highly inefficient in these objects \citep{penarrubia2012}.  This is a particular problem at the low-mass end of the Galactic satellite population; measuring the central dark matter profiles of the lowest-mass objects is vital.

Addressing such issues using dynamical modelling requires that both the data and the models themselves are good enough to obtain meaningful answers.  In the past, the lack of conclusive results was due to the amount and the quality of the data that was available; small datasets with large errors restricted the type of dynamical modelling that could be done.  For some systems, the data is still the limiting factor, but this is not always the case.  For objects in the Local Group - that is our own Milky Way, sister galaxy Andromeda (M31) and their globular clusters and dwarf galaxy satellites - we are in the fortunate position of being able to measure photometric and spectroscopic quantities for individual stars, often to very high precision, thanks to both their proximity and advances in modern observing techniques.  These data often include not only line-of-sight velocities, but motions on the plane of the sky (proper motions) and metal abundances.  Combining the line-of-sight velocities and proper motions gives us the full 3-dimensional velocities of the stars, instead of projections of that velocity.  From this we can directly calculate the velocity anisotropy and, thus, break the degeneracies that exist between the velocity anisotropy of the stars and the shape of the potential in which they orbit.

The \textit{Hubble Space Telescope} (\textit{HST}) has delivered data of exceptional accuracy, including proper motions for globular clusters \citep[e.g.][]{anderson2010, bellini2013}.  Recent studies using ground-based data have proved that they too can produce remarkable results \citep[e.g.][]{bellini2009}; such datasets are complementary as they probe a much larger radial extent than is feasible with the \textit{HST}.  Large-scale surveys may lack the sensitivity of the \textit{HST} but are a vital tool due to the sheer number of stars that they have observed; recent surveys such as HIPPARCOS \citep{esa1997}, the Sloan Digital Sky Survey \citep[SDSS;][]{york2000} and the RAdial Velocity Experiment \citep[RAVE;][]{steinmetz2006} have provided a wealth information over large regions of the sky, which have been used to good effect in studies of the Milky Way.  For studying the smaller denizens of the Local Group, there have also been a number of focused observing efforts; for example, \citet{walker2009} have published data for four of the Milky Way's classical dwarfs: Carina, Fornax, Sculptor and Sextans.  And the future is bright: there are a number of surveys coming online over the next few years that will expand the data sets that are currently available.  For the Milky Way in particular, Gaia will provide velocities and abundance measurements of unprecedented accuracy over the whole sky \citep{perryman2001}.  These data are nothing without the tools to properly analyse them and we must ensure we have those tools in place both to analyse the existing data and to fully exploit the promised data when they become available.

Current dynamical modelling techniques often do not do justice to such datasets; typically, they proceed by spatially binning the data and comparing the velocity moments in each of the bins with the velocity moments predicted by a theoretical model.  When calculating velocity moments for a set of data, we necessarily assume that all stars are members of the object that we are modelling.  If we suspect that our data set contains contaminants, then we must take care to remove them before binning.  This is a tricky endeavour as the member and contaminating velocity distributions often overlap and it is difficult to disentangle the two populations.  Being too conservative with membership cuts will excise true members; being too lenient will retain non-members.  In either case, the resulting velocity moments will be estimated incorrectly.  Membership selection aside, binning methods suffer from a loss of information.  In order to estimate velocity moments, each bin must contain a sufficient number of stars.  For the first and second moments - means and dispersions respectively - an average of 50 stars per bin is usually enough (for higher velocity moments, the number of stars required increases).  Even in the simplest case, a dataset of a few thousand stars will be reduced to only a few tens of bins.  Finally, the comparison of estimated and model moments is often done using simple $\chi^2$ techniques.

Binning approaches are clearly flawed, so can we do better?  Yes we can.  We have developed existing dynamical modelling techniques to directly fit discrete data using maximum likelihood methods.  Fitting each star individually means that we no longer have to make any quality cuts on our datasets; we can simple include a contaminating population in our models and fit for that too.  Another advantage of using maximum likelihood methods to fit the kinematics, is that the likelihoods can be extended easily to incorporate further information, such as metal abundances.  This particular application goes beyond the scope of this paper, but the analysis we present here is readily extensible in such a fashion.

One prime example that highlights all we are currently able to achieve is the Galactic globular cluster \wcen{} (NGC\,5139).  Located only 5~kpc from the Sun, it is large and bright and has been observed many times with many different instruments, over a long time baseline.  As a result, there are line-of-sight velocities and abundances available for thousands of stars \citep[e.g.][]{suntzeff1996, mayor1997, reijns2006, pancino2007} and proper motions measurements available for hundreds of thousands of stars \cite[e.g.][]{vanleeuwen2000, bellini2009, anderson2010}.  \wcen{} is an interesting object to study as it demonstrates many qualities in common with globular clusters, such as an apparent absence of dark matter, and also many qualities in common with dwarf spheroidal galaxies, such as a complex star formation history.  There is also an ongoing debate concerning the presence (or absence) of an intermediate-mass black hole (IMBH) at its centre \citep{noyola2008, vandermarel2010, noyola2010}.

With so many unanswered questions surrounding its structure and origins, \wcen{} has been the focus of many studies stretching back over many years.  One such study by \citet{vandeven2006} used both line-of-sight velocity and proper motion data to perform a (binned) axisymmetric \citet{schwarzschild1979} analysis of the cluster from which they were able to constrain the cluster distance, inclination, mass-to-light ratio and mass to good accuracy.  The large size and high quality of the datasets they used, combined with the powerful orbit-based models, showed that analysis of binned data can be very effective.  However, smaller datasets will suffer from the binning process and will not be so successful.

Our ultimate goal is to develop discrete models that use sophisticated modelling techniques -- such as Schwarzschild's orbital-superposition method or made-to-measure methods \citep[e.g.][]{syer1996, delorenzi2007, long2010} -- that are able to handle the physical complexities of dynamical systems \citep[see also][]{wu2006, wu2007, chaname2008}.  By eliminating the need for binning and working directly with discrete data, both data and models will be able to reach their full potential.  However, at present, we are using a Jeans' analysis to do the modelling as they are simpler to understand and computationally less expensive than the alternatives.  The particular class of models that we use are not ideal: they assume that the velocity ellipsoid is aligned with the cylindrical coordinate system; they fix ab initio the relative contributions of ordered and random motions; and they cannot exclude unphysical (negative) distribution functions.  Despite these disadvantages, they are adequate for our purposes while we are developing our machinery and working to understand the data quality we require.

Here we present the first steps towards developing discrete dynamical modelling techniques, which we apply to data for \wcen{}.  As our goal here is to test the power of our likelihood formalism, we use the existing dataset from \citet{vandeven2006} that has been rigorously tested and well-studied.  \autoref{sect:data} describes the photometric and kinematic data for \wcen{}, \autoref{sect:maxlh} develops the maximum-likelihood analysis and \autoref{sect:jam} outlines the cluster models.  In \autoref{sect:wcen}, we apply our methods to \wcen{}.  We discuss our results in \autoref{sect:discuss} and conclude in \autoref{sect:conc}.  In a companion paper, we use a similar analysis to study the mass-to-light profile of globular cluster M15 \citep{denbrok2013}.

% -----------------------------------------------------------------------------

\section{Data}
\label{sect:data}

The dynamical models we will use require a surface brightness profile for \wcen{} in the form of a Multi-Gaussian Expansion \citep[MGE][]{emsellem1994}; we adopt the MGE derived by \citet{dsouza2013}.  For the kinematics, we use the same proper motion and line-of-sight velocity datasets and reduction techniques described in Sections 2-4 of \citet{vandeven2006}.  We briefly introduce the MGE and outline the steps taken to combine and clean the kinematic datasets, but refer the reader to the original papers for more complete descriptions.

\subsection{Multi-gaussian expansion}

\begin{figure*}
\begin{center}
    \includegraphics[width=0.47\linewidth]{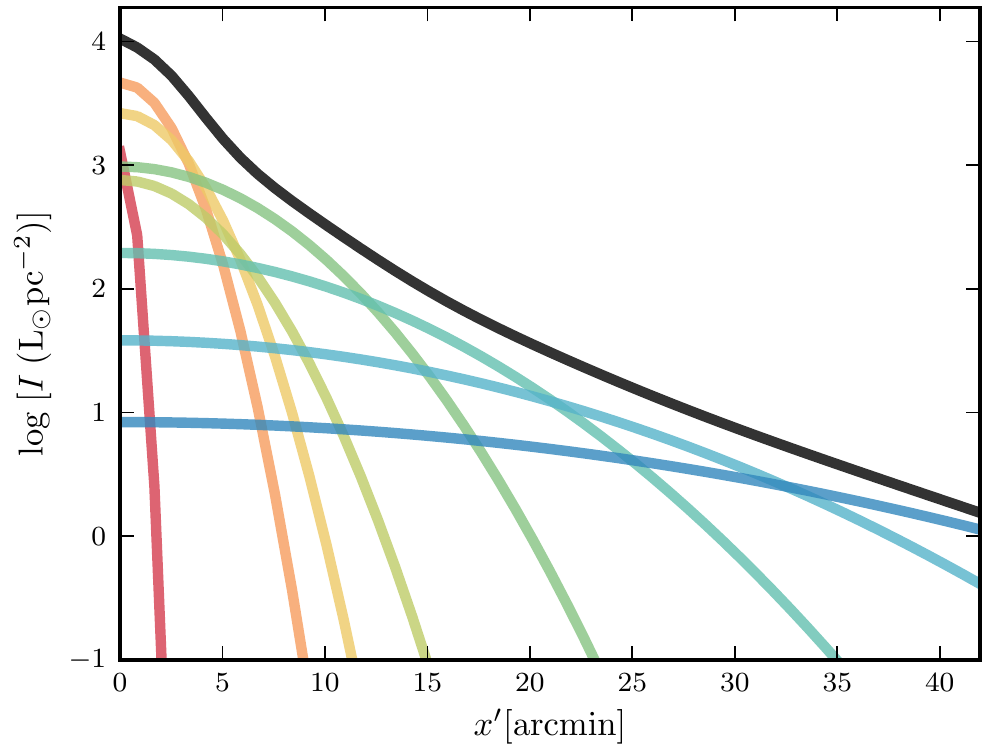}
    \qquad
    \includegraphics[width=0.47\linewidth]{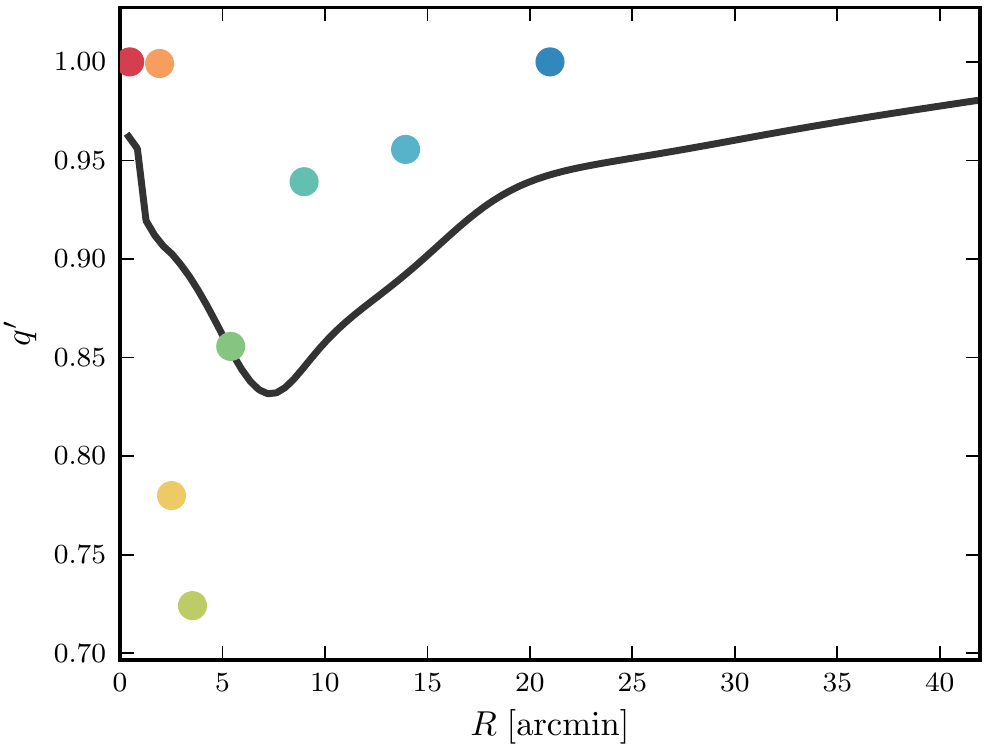}
    \caption{\textit{Left:} The surface brightness profile of the adopted MGE model along the projected major axis.  The black line traces the total surface brightness.  The coloured lines trace the surface brightness profiles of the individual gaussian components.  \textit{Right:} The ellipticity profile of the adopted MGE model (black line).  The coloured points show the flattenings of the individual MGE components at their corresponding dispersions.  The colours identify the same components in both panels.}
    \label{fig:mge}
\end{center}
\end{figure*}

\begin{table}
    \caption{Multi-gaussian expansion for \wcen{} from \citet{dsouza2013}. For each gaussian component: $k$ is the gaussian number; $L_k$ is the central surface brightness; $\sigma_k$ is the dispersion along the major axis; $q'_k$ is the projected flattening.  $\kappak$ is the rotation parameter derived from kinematic fits.}
    \label{table:mge}
    \begin{center}
    \begin{tabular}{ccccc}
        \hline
        \hline
        $k$ & $L_k$ & $\sigma_k$ & $q'_k$ & $\kappa_k {}^a$ \\
        & ($\mathrm{L}_{\odot}\,\mathrm{pc}^{-2}$) & (arcmin) & \\
        \hline
        1 & 1290.195 & 0.475570 & 1.0000000 &  0.0 \\
        2 & 4662.587 & 1.931431 & 0.9991714 &  0.0 \\
        3 & 2637.784 & 2.513385 & 0.7799464 & -0.4 \\
        4 & 759.8591 & 3.536726 & 0.7241260 & -1.1 \\
        5 & 976.0853 & 5.403728 & 0.8556435 & -0.6 \\
        6 & 195.4156 & 8.983056 & 0.9392021 &  0.0 \\
        7 & 38.40327 & 13.93625 & 0.9555874 &  0.0 \\
        8 & 8.387379 & 20.98209 & 1.0000000 &  0.0 \\
        \hline
        \hline
    \end{tabular}
    
    \raggedright
    \medskip
    $^a$We have negated the non-zero rotation parameters as we use a position angle of -80$^\circ$ for the major axis, where \citet{dsouza2013} used a position angle of 100$^\circ$.
    \end{center}
\end{table}

MGE models provide a smooth representation of the surface brightness profile of a stellar system; they are formed from the sum of a set of gaussian components, each of which is defined by a central surface brightness, a major-axis dispersion and a flattening.  Formulae relating to MGE surface brightness expansions, mass density and potential are given in \autoref{sect:jamapp}.  Gaussians do not provide a complete set of functions and, thus, cannot provide an exact fit to a surface brightness profile; nevertheless, in most cases, the profiles are accurately reproduced.

The MGE from \citet{dsouza2013} that we use throughout this paper is given in the first four columns of \autoref{table:mge}.  This was derived by fitting to the radial number density profile of \wcen{} from \citet{ferraro2006} and optimised to simultaneously reproduce the flattening profile from \citet{geyer1983}.  In the left-hand panel of \autoref{fig:mge}, we show surface brightness profile of the MGE along the major axis; the total profile is shown in black and the profiles for the individual gaussian components are shown as coloured lines.  In the right-hand panel of \autoref{fig:mge}, we show the ellipticity profile of the MGE (black line); the coloured points show the flattenings of the individual gaussian components.

\begin{table*}
    \caption{The six data subsets we use for our analysis.  Dataset A is the cleaned sample that contains only members of \wcen{}, achieved by applying a series of quality cuts.  Datasets B-E gradually relax each of the cuts in turn.  Dataset F is the full data set with no cuts applied.  The table presents the line-of-sight velocity (LV) cuts and proper-motion (PM) cuts applied, the number of stars in the dataset and a brief description of the dataset.}
    \label{table:datasets}
    \begin{center}
    \begin{tabular}{ccccl}
        \hline
        \hline
        dataset & LV cuts & PM cuts & stars & description \\
        \hline
        A & i, ii & i, ii, iii, iv, v & 3740 & ``clean'' dataset, identical to the final sample used by \citet{vandeven2006}. \\
        B & - & i, ii, iii, iv, v & 4655 & full LV dataset; all PM cuts still in place. \\
        C & - & iv, v & 4851 & includes suspected PM non-members but continues to cut out blended and high-error PM stars. \\
        D & - & v & 5220 & includes blended stars (classes 0-3) but keeps the cut on magnitude of error. \\
        E & - & iv & 7371 & includes high-error stars but keeps the cut on blending. \\
        F & - & - & 9511 & no cuts; contains all LV and PM stars. \\
        \hline
        \hline
    \end{tabular}
    \end{center}
\end{table*}

\subsection{Kinematic data}
\label{sect:data_process}

The proper motions come from the catalogue of \citet{vanleeuwen2000}, which contains 9847 stars.  We rotate the positions and proper motions through a position angle of -80$^{\circ}$ in order to align the coordinate axes on the plane of the sky ($x'$, $y'$) with the major and minor axes respectively.  Following \citet{vandeven2006}, we further correct the proper motions for perspective rotation -- using a distance of $D = 5.0 \pm 0.2$~kpc \citep{harris1996} and systemic velocities $\mu_{\alpha}^{\mathrm{sys}} = -3.97 \pm 0.41$~\masyr, $\mu_{\delta}^{\mathrm{sys}} = -4.38 \pm 0.41$~\masyr \citep[both][]{vanleeuwen2000} and $\mu_{\mathrm{LOS}}^{\mathrm{sys}} = 232$~\kms \citep{reijns2006} -- and for solid body rotation of $\Omega = 0.029$ \masyr\,arcmin$^{-1}$.  Finally, we adjust the proper motions for any residual systemic velocity components, which are calculated using only ``clean'' stars (see \autoref{sect:data_select}).

In crowded fields, such as we have for \wcen{}, stars may overlap and appear to be blended, thus decreasing the accuracy with which stellar positions -- and, hence, proper motions -- can be measured.  Each star in the proper motion catalogue is assigned a class flag based on its distance from other stars, where class 0 stars are unaffected by nearby stars and class 4 stars are badly blended.  We will use this class flag later to select a subset of the proper motion stars that we are sure are unblended.  The catalogue also offers a membership probability for each star as a percentage, which we use later to select only likely members.

For the line-of-sight velocities, we start with data for 360 stars from \citet{suntzeff1996}, 471 stars from \citet{mayor1997}, 1966 stars from \citet{reijns2006} and 4916 stars kindly provided by Karl Gebhardt (private communication).    These datasets overlap so some stars appear more than once (though no star appears in every dataset).  These datasets are first crossmatched using catalogue identification numbers, where they exist.  We discard the nine stars remaining without a position measurement and a further seven stars without a positive velocity error measurement.  After ensuring that all the datasets are aligned with the coordinate system used for the proper motion catalogue, we again rotate the positions through a position angle of -80$^{\circ}$ in order to align the major and minor axes.  Following advice from Gebhardt (private communication), we discard a further 3564 stars fainter than 14.5~mag at this stage as some stars were smeared during the reduction process and misidentified as two fainter stars.  Finally, we combine the velocities of any stars appearing in more than one of the four datasets.  The final line-of-sight velocity dataset contains 3094 unique stars.

\subsection{Kinematic samples}
\label{sect:data_select}

In \citet{vandeven2006}, the datasets were thinned down further using a series of cuts to ensure that only bona fide cluster members with accurate velocities were included in the analysis.  This was particularly important for their study as they binned their final dataset and calculated velocity moments in each bin; non-member contaminants and large velocity uncertainties would have led to systematic errors in the calculated moments.

The aim of our study is to develop a method that is able to handle all contaminating populations and large uncertainties, however, we wish to test our method first on ``clean'' data.  We use the same criteria as \citet{vandeven2006} to select a clean sample.  For the line-of-sight velocity (LV) dataset, a star is retained if:
\begin{enumerate}
    \item the uncertainty on the line-of-sight velocity is smaller than 2.0~\kms;
    \item the uncorrected line-of-sight velocity lies in the range $160 < \vlos < 300$~\kms
\end{enumerate}
For the proper motion (PM) data set, a star is retained if:
\begin{enumerate}
    \item the star is known to be a cluster member from cross-matching with the line-of-sight velocity dataset (3762 stars);
    \item the probability of membership is greater than 68\% (only for stars not in the line-of-sight velocity dataset);
    \item the proper motions are within 5$\sigma$ (3.6~\masyr in right ascension and 3.2~\masyr in declination) of the cluster velocity peak;
    \item the star is unblended (class 0);
    \item the average uncertainty on the proper motions is smaller than 0.2~\masyr.
\end{enumerate}
The cleaned line-of-sight velocity dataset contains 2163 stars and the cleaned proper motion dataset contains 2295 stars.  There is some overlap between the datasets; the clean combined dataset contains 3740 unique stars.

Once we have verified that our methods work with this ``clean'' data set, we will relax the cuts in a series of steps to see how well our models deal with contaminants and stars with high velocity errors.  We start by relaxing the cuts on the line-of-sight velocity dataset first and then move onto the proper-motion cuts.  The kinematics subsamples we will use are described in \autoref{table:datasets}.

% -----------------------------------------------------------------------------

\section{Maximum likelihood analysis}
\label{sect:maxlh}

Our goal is to model a discrete dataset without first binning the data; instead we will compare models against our dataset on a star-by-star basis using a maximum-likelihood analysis.  This will also allow us to include contaminants instead of making cuts on the data to remove them.  Here we describe the principles of a maximum-likelihood analysis, the inclusion of a contaminating population in our models, and the form of the cluster models and contamination models that we will use for our study of \wcen{}.

\subsection{Likelihood}

Consider a dataset of $N$ stars such that the $i$th star has coordinates $\bxi = (x'_i, y'_i)$ and velocities $\vxi \pm \svxi$, $\vyi \pm \svyi$ and $\vzi \pm \svzi$, where $x'$ is the direction of the projected major axis, $y'$ is the direction of the projected minor axis and $z'$ is the direction along the line of sight. The velocity vector $\vi$ is then
\begin{equation}
    \vi = \left( \vxi, \vyi, \vzi \right)
\end{equation}
and the error matrix $\Si$ is
\begin{equation}
    \Si = \left( \begin{array}{ccc}
        \svxi^2 & 0 & 0 \\
        0 & \svyi^2 & 0 \\
        0 & 0 & \svzi^2
    \end{array} \right)
\end{equation}
if we assume that the measurements are uncorrelated.

Suppose we have a set of models and we wish to know which model is able to best describe the dataset.  Let $\bTheta_j$ represent the parameter set for a particular model; then the likelihood of observing star $i$ given model $j$ is given by
\begin{equation}
    \Lij = p \left( \vi \, \left| \, \bxi, \Si, \bTheta_j \right. \right)
\end{equation}
Note that we are treating the position data as prior information, so this likelihood is a probability distribution function for the velocities only and depends upon the position of the star.  We adopt this approach because selection effects can be complicated, which means we are unable to model the number density distribution of the cluster.  Instead we wish to model the velocity distribution of the cluster given the positions of the stars that we have.

Now the total likelihood $\Lj$ of the model is product of the model likelihoods for each star is
\begin{equation}
    \Lj = \prod_{i=1}^N \Lij .
\end{equation}
In practice, it is often easier to work with log-likelihoods $\lj = \ln \Lj$.  The best model is the set of parameters $\bTheta_j$ that maximises $\Lj$ and, hence, $\lj$.

\subsection{Contaminants}
\label{sect:contams}

Now further suppose our dataset contains a contaminating foreground or background population in addition to stars belonging to the cluster under study.  Then parameter set $\bTheta_j$ comprises a model for the cluster with parameters $\bThcl$, which has a likelihood $\Lijcl$, and a model for the background population with parameters $\bThbg$, which has a likelihood $\Lijbg$.  Then the likelihood becomes
\begin{equation}
    \Lij = \eta_i \, \Lijcl + \left( 1 - \eta_i \right) \, \Lijbg
\end{equation}
where $\eta_i = 1$ if the star is a member of the cluster and $\eta_i = 0$ if the star is part of the contaminating population (intermediate values are not permitted).  Unfortunately, we do not know which stars are cluster members and which stars are contaminants, so the $\eta_i$ values are unknown.  This adds $N$ extra parameters to $\bTheta_j$, which is unfeasible for modelling purposes.

Instead we introduce a mixture model that combines the cluster and background likelihoods
\begin{equation}
    \Lij = \pmi \left( \bxi \right) \, \Lijcl
        + \left[ 1  - \pmi \left( \bxi \right) \right] \, \Lijbg
\end{equation}
where $\pmi \left( \bxi \right) = p \left( \eta_i = 1 \, \left| \, \bxi \right. \right)$ is the prior probability that the star is a member of the cluster given its position, and hence $1 - \pmi \left( \bxi \right)$ is the prior probability that the star is a member of the background population given its position.

Note, the posterior membership probability for each star under a given model can be calculated via
\begin{equation}
    m_{i} = p \left( \eta_i = 1 \right) = \frac{\pmi \Lijcl}{\pmi \Lijcl + \left( 1 - \pmi \right) \Lijbg} .
\end{equation}
These membership probabilities will be particularly useful once a best model has been found.

\subsection{Membership priors}

Parameter $\pmi$ is the prior probability that star $i$ is a member of the cluster, given certain of its observed properties.  In this paper, we will use only the position of a star to determine the prior on its membership probability, however this method can be extended to also account for other properties -- such as magnitudes, colours and metallicities -- when considering whether a star is likely to be a member of the cluster or the background.

Stars near the projected centre of the cluster are more likely to be cluster members than stars in the outer parts.  The MGE model for the cluster gives us the luminosity surface density $I$ of the cluster, as given by \autoref{eqn:mgesurf}.  Our dataset contains only red giant stars; these have a narrow range of magnitudes, thus we assume that there is a constant factor that will allow us to convert luminosity surface density $I$ (in $\Lsun \mathrm{pc}^{-2}$) into number surface density $\mathrm{d} N_\mathrm{cl}$ (in $\mathrm{arcsec}^{-2}$) such that
\begin{equation}
    \df N_\mathrm{cl} \left( \bxi \right) \propto I \left( \bxi \right) .
\end{equation}
We also assume that there is a background number surface density $\mathrm{d} N_\mathrm{bg}$.  Then the prior probability of cluster membership is given by
\begin{equation}
    \pmi \left( \bxi \right) = \frac{\df N_\mathrm{cl} \left( \bxi \right)}{\df N_\mathrm{cl} \left( \bxi \right) + \mathrm{d} N_\mathrm{bg} \left( \bxi \right)}
\end{equation}
The background number density contribution is unknown.  We assume that it is constant throughout the observed cluster field and is equal to some fraction $\epsilon$ of the central cluster number surface density $\mathrm{d} N_0 = \mathrm{d} N_\mathrm{cl} \left( 0, 0 \right)$. Then the prior on cluster membership becomes
\begin{equation}
    \pmi \left( \bxi \right) = \frac{\df N_\mathrm{cl} \left( \bxi \right)}{\df
        N_\mathrm{cl} \left( \bxi \right) + \epsilon \, \mathrm{d} N_0}
        \label{eqn:postmem}
\end{equation}
where $\epsilon$ is unknown and will be a free parameter in our models.

\subsection{Cluster and background models}

\begin{figure}
\begin{center}
    \includegraphics[width=\linewidth]{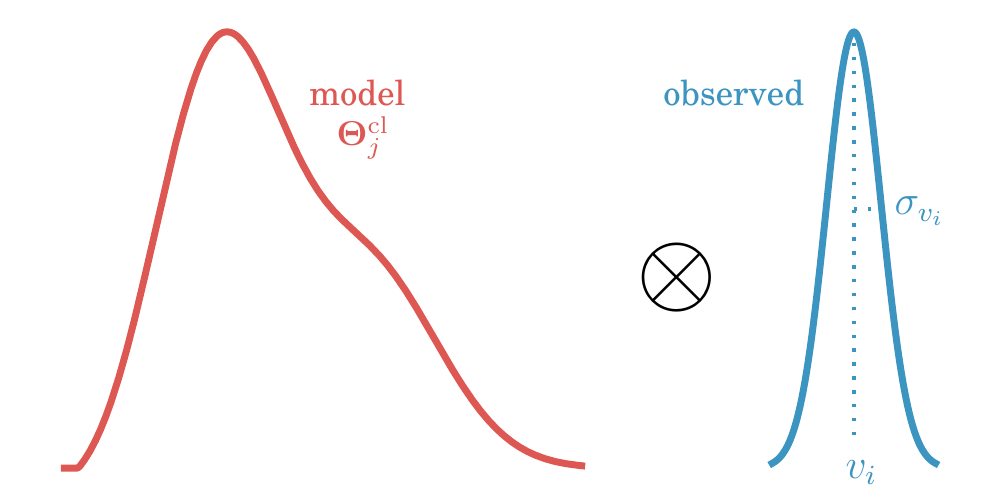}
    \caption{Cartoon.  The red curve (left) shows the velocity distribution generated by a (cluster) model with parameters $\bThcl$ at the position of the $i$th star.  The blue curve (right) shows a gaussian with a mean equal to the observed velocity $v_i$ of the $i$th star and a width equal to the uncertainty $\sigma_{v_i}$ on the observed velocity.  In order to determine the likelihood of the observed velocity given the measurement uncertainty and model predictions $\Lijcl$, we convolve these two distributions.}
    \label{fig:veldbncartoon}
\end{center}
\end{figure}

Likelihood $\Lijcl$ is the probability of the observed velocity $\vi$ given the measurement uncertainties $\Si$ and the cluster-velocity distribution predicted by the model parameters $\bThcl$ at the position of the star, which can be expressed as
\begin{equation}
    \Lijcl = p \left( \vi \, \left| \, \bxi, \Si, \eta_i = 1, \bThcl \right.
        \right) .
\end{equation}
In practice, we convolve the velocity distribution predicted by the cluster model with a gaussian distribution representing the observed velocity and its uncertainty.  We illustrate the one-dimensional case by the cartoon in \autoref{fig:veldbncartoon}.  The red curve (left) shows the velocity distribution generated by a model with parameters $\bThcl$ at the position of the $i$th star.  The blue curve (right) shows a gaussian with a mean equal to the observed velocity $v_i$ and a width equal to the uncertainty $\sigma_{v_i}$ on the observed velocity.  The model predictions will change with position and, thus, will be different for each star.

We will proceed by assuming that the velocity distribution predicted by the model is a tri-variate gaussian with mean velocity $\bmucl$ and covariance $\bCcl$ at $\bxi$.  The likelihood becomes
\begin{align}
    \Lijcl & = p \left( \vi \, | \, \bxi, \Si, \eta_i = 1, \bmucl, \bCcl
        \right) \nonumber \\
    & = \frac{ \exp \left[ - \frac{1}{2} \left( \vi - \bmucl
        \right)^{\mathrm{T}} \left( \bCcl + \Si \right)^{-1} \left( \vi -
        \bmucl \right) \right] } { \sqrt{ \left( 2 \pi \right)^{n} \left|
        \left( \bCcl + \Si \right) \right| } }
    \label{eqn:lijcl}
\end{align}
where $n$ is the rank of $\Si$; as our hypothetical dataset has proper motions and line-of-sight velocities, $n = 3$.  Similarly, we assume that the background model predicts a multi-variate-gaussian velocity distribution with mean $\bmubg$ and covariance $\bCbg$.

Thus far, we have considered a dataset where all stars have full velocity information, but this is not true for our \wcen{} dataset.  Only a subset of the stars have all three velocity components, a further subset have only proper-motion measurements, and the rest have only a line-of-sight velocity measurement.  However, we can still use the same modelling analysis; for stars with only 1 or 2 velocity components we simply use a uni-variate $(n = 1)$ or bi-variate $(n = 2)$ gaussian for the cluster and background models.

% ---------------------------------------------------------------------------- %

\section{Jeans' Models}
\label{sect:jam}

We model the cluster using an extended version of the axisymmetric Jeans Anisotropic MGE (JAM) formalism described by \citet{cappellari2008}, who presented the projected first- and second-moment calculations for line-of-sight velocities only.  Recently, \citet{dsouza2013} presented second-moment calculations for the major- and minor-axis proper motions, and \citet{cappellari2012} presented the second moment cross-terms.  However, the first moments for the major- and minor-axis proper motions remain uncalculated, so we do so here.  For completeness, we include a derivation of all the first- and second-moment equations.  The calculations are given in full in \autoref{sect:jamapp}, we present only a brief introduction and the final equations here.

We work within an axisymmetric framework where cylindrical polar coordinates $(R, \phi, z)$, with $R^2 = x^2 + y^2$, describe the intrinsic coordinates of the system and $(x', y', z')$ are the projected coordinates on the plane of the sky; the $x'$-axis is aligned with the projected major axis, the $y'$-axis with the projected minor axis, and the $z'$-axis lies along the line-of-sight such that the line-of-sight vector is positive in the direction away from us.

For an axisymmetric ($\frac{\partial}{\partial \phi} = 0$) system in a steady state ($\frac{\partial}{\partial t} = 0$), the second moment Jeans equations in cylindrical polars are
\begin{align}
    \frac{\nu (\vRsqbar - \vphisqbar)}{R}
        + \frac{\partial (\nu \vRsqbar)}{\partial R}
        + \frac{\partial (\nu \vRvzbar)}{\partial z}
        & = - \nu \frac{\partial \Phi}{\partial R} \\
    \frac{\nu \vRvzbar}{R}
        + \frac{\partial (\nu \vRvzbar)}{\partial R}
        + \frac{\partial (\nu \vzsqbar)}{\partial z}
        & = - \nu \frac{\partial \Phi}{\partial z}
\end{align}

In order to obtain a unique solution for the second moments from these equations, we make two assumptions: that the velocity ellipsoid is aligned with cylindrical polar coordinate system, so that $\vRvzbar = 0$; and that the anisotropy is constant and quantified by $\vRsqbar = b \vzsqbar$.  The projected second velocity moments along the line of sight are then given by
\begin{align}
    I \, \overline{v_{\alpha} v_{\beta}} \left( x',y' \right) = {} & \fpigu
        \nqpui \nonumber \\
    {} & \times \vvdenom \nonumber \\
    {} & \times \expabxy \df u
\end{align}
where $\alpha$ and $\beta$ are the projected coordinate directions $x'$, $y'$ and $z'$ and $\iabk$ for each moment is
\begin{align}
    \ii_{x' x', k} & = b_k \sksq q_k^2 + \frac{\dd}{\ee^2} \ab^2 y'^2 \cos^2 i
        + \frac{\dd}{2 \ee} \sin^2 i \\
    \ii_{y' y', k} & = \left( b_k \cos^2 i + \sin^2 i \right) \sksq q_k^2
        + \dd x'^2 \cos^2 i \\
    \ii_{z' z', k} & = \left( b_k \sin^2 i + \cos^2 i \right) \sksq q_k^2
        + \dd x'^2 \sin^2 i \\
    \ii_{x' y', k} & = - \frac{\dd}{\ee} x' y' \cos^2 i \ab \\
    \ii_{x' z', k} & = \frac{\dd}{\ee} x' y' \cos i \sin i \ab \\
    \ii_{y' z', k} & = \left[ (1 - b_k) \sksq q_k^2 - \dd x'^2 \right]
        \cos i \sin i .
\end{align}
The subscripts $j$ and $k$ refer to quantities associated with the potential and luminous gaussians respectively.

To calculate the first moments, we must make a further assumption in order to obtain a unique solution: we set the relative contributions of random and ordered motion to the RMS velocities via a rotation parameter $\kappak$ for each component of the luminous MGE such that
\begin{equation}
    \nuvphik = \kappak \left( \nuvphisqk - \nuvrsqk \right)^{\frac{1}{2}} .
    \label{eqn:kappak}
\end{equation}
The projected first velocity moments along the line of sight are given by
\begin{equation}
    I \, \overline{v_{\tau}} \left( x',y' \right) = 2 \sqrt{\pi G} \intinf
        \ff_{\tau} \gg \df z'
\end{equation}
where $\tau$ represents the projected coordinate directions $x'$, $y'$ and $z'$ and $\ff_{\tau}$ for each moment is
\begin{align}
    \ff_{x'} & = y' \cos i - z' \sin i \\
    \ff_{y'} & = x' \cos i \\
    \ff_{z'} & = x' \sin i .
\end{align}
All further terms are defined in \autoref{sect:jamapp}.

\citet{cappellari2008} kindly made available his IDL code.  We have converted this code into C in order to speed up the run time and extended it to calculate all three first moments and all six second moments.  This code is available from http://github.com/lauralwatkins/cjam.

The maximum likelihood formalism described in \autoref{sect:maxlh} requires the velocity vector and covariance matrix at each point $(x',y')$.  These are easily obtained from the first and second moments.  The velocity prediction $\bmu$ for the model is given by
\begin{equation}
    \bmu = \left( \vxm, \vym, \vzm \right)
\end{equation}
and the covariance matrix $\bC$ for the model is given by
\begin{equation}
    \bC = \left( \begin{array}{ccc}
        \vsqxm - \vxm^2 & \vsqxym - \vxm\,\vym & \vsqxzm - \vxm\,\vzm \\
        \vsqxym - \vxm\,\vym & \vsqym - \vym^2 & \vsqyzm - \vym\,\vzm \\
        \vsqxzm - \vxm\,\vzm & \vsqyzm - \vym\,\vzm & \vsqzm - \vzm^2
    \end{array} \right) .
\end{equation}

% ---------------------------------------------------------------------------- %

\section{Application to $\omega$ Centauri}
\label{sect:wcen}

Now that we have laid out our discrete maximum likelihood methods, we can apply them to \wcen{}.  We fix the cluster systemic velocity at $\mu^{\mathrm{sys}}_{x'} = 3.88 \pm 0.41 \; \mathrm{mas \; yr^{-1}}$, $\mu^{\mathrm{sys}}_{y'} = -4.44 \pm 0.41 \; \mathrm{mas \; yr^{-1}}$, $v^{\mathrm{sys}}_{z'} = 232.02 \pm 0.03 \; \mathrm{km \; s^{-1}}$ \citep{vandeven2006}.  The contaminants will predominantly be dwarf stars in the Milky Way so the contaminating population in our models is assumed to have a velocity of $-\mathbf{v}_{\mathrm{sys}}$ with a dispersion\footnote{We choose a dispersion typical of Milky Way dwarf stars, although we note that our analysis is not sensitive to this value.} of 50~\kms{} in all directions, which corresponds to $\sim$0.2~\masyr{} at the assumed distance of \wcen{}.

\citet{dsouza2013} determined rotation parameters\footnote{These parameterise the proportions of ordered and random motions via \autoref{eqn:kappak}, effectively setting the amount of rotation for the gaussian.} $\kappak$ for their \wcen{} MGE; these values are given in the final column of \autoref{table:mge}.  They were derived by fitting to the cleaned proper motion data set from \citet{vandeven2006}, the same proper motion dataset that we use in datasets A and B.  The values of the rotation parameters vary from $\kappa = 0.$ to $\kappa = -1.1$, so it clear that adopting a single value of $\kappa$ for the whole cluster would be incorrect.  However, fitting the rotation parameters for the cluster is beyond the scope of this paper; thus, we fix the rotation parameters to the \citet{dsouza2013} values.

We are left with five free parameters in our models:
\begin{enumerate}
    \item $\lambda = - \ln \left( 1 - \beta \right)$, where $\beta$ is the velocity anisotropy and assumed constant for the system, hence $b_k = \frac{1}{1 - \beta}$ for all $k$;
    \item $\Upsilon$, the mass-to-light ratio, which we assume is constant for the system;
    \item $\qbar$, the median intrinsic flattening of the MGE, which is related to the inclination angle $i$ via $\cos i = \sqrt{ \frac{\overline{q'}^2 - \qbar^2}{1 - \qbar^2} }$ where $\overline{q'}$ is the median projected flattening of the MGE;
    \item $d$, the distance;
    \item $\epsilon$, the contamination fraction.
\end{enumerate}
We expect that the system is approximately isotropic, thus $\beta \sim 0$ and, hence, $\lambda \sim 0$.  \citet{vandeven2006} found an inclination angle $i = 50^{\circ} \pm 4^{\circ}$, which gives median flattening $\qbar \sim 0.91$ for the MGE we have adopted, and a distance $d = 4.8 \pm 0.3$~kpc.  They also found a V-band mass-to-light ratio $\Upsilon = 2.5 \pm 0.1 \; \Msun / \Lsun$, however \citet{dsouza2013} showed that the MGE calculated by \citet{vandeven2006} is inconsistent with the projected flattening of \citet{geyer1983}; as a result their mass and, hence, mass-to-light ratio is underestimated by $\sim$7\%.  By using an improved MGE fit, we expect to find a mass-to-light ratio of $\sim$2.7.  Contamination fraction $\epsilon$ will, of course, depend upon the dataset that we use: for the cleaned dataset A where we retain only member stars, we expect that $\epsilon \sim 0$; as we relax the cuts, we expect that $\epsilon$ will increase.  Here we have considered only the parameters estimated by the \citet{vandeven2006} study because we are using the same datasets and so expect to obtain similar results if our models are successful.  We note that these values are generally representative of the set of literature estimates, however we will return to this point in \autoref{sect:bestmodel}.

In order to efficiently sample our parameter space, we turn to Markov Chain Monte Carlo analysis; we use the \textsc{emcee} package developed by \citet{foremanmackey2013}, which is an implementation of the affine-invariant MCMC ensemble sampler by \citet{goodman2010}. At each step, the algorithm uses a set of walkers to explore the parameter space; the results from each of the walkers informs the next choice of models to be evaluated.  We run our models with 100 walkers.

As we have already discussed, the goal of this paper is to develop discrete dynamical modelling tools that are able to deal with contaminants.  However, to first test our models, we apply them to a cleaned dataset (dataset A) for which we are confident that nearly all stars are members of \wcen{}.  Then we relax the membership and quality cuts and investigate the performance of the models when contaminants are present (datasets B-F).  The six datasets we use are described in \autoref{sect:data_select}.

\subsection{Cleaned data set}

\begin{figure}
\begin{center}
    \includegraphics[width=\linewidth]{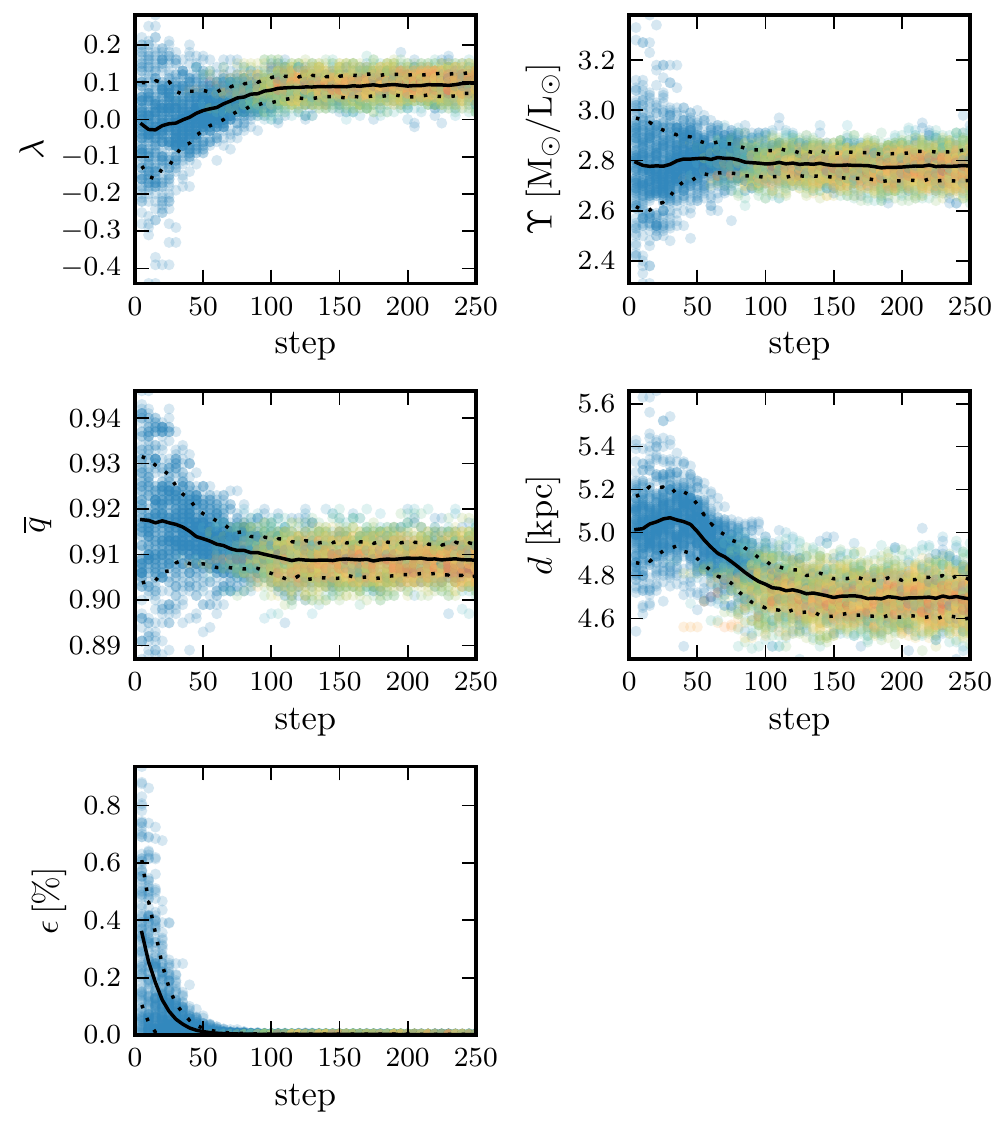}
    \caption{MCMC chain evolution for dataset A.  Top left: anisotropy parameter $\lambda = - \ln \left( 1 - \beta \right)$.  Top right: V-band mass-to-light ratio $\Upsilon$.  Middle left: median intrinsic flattening $\qbar$.  Middle right: distance $d$.  Bottom left: contamination fraction $\epsilon$.  The points show the values visited by the walkers at each step, coloured by their likelihood from red (high) to blue (low).  The solid (dotted) lines show the means (dispersions) of the walker values at each step.  The MCMC chain converges tightly.}
    \label{fig:seta_evol}
\end{center}
\end{figure}

\begin{figure*}
\begin{center}
    \includegraphics[width=\linewidth]{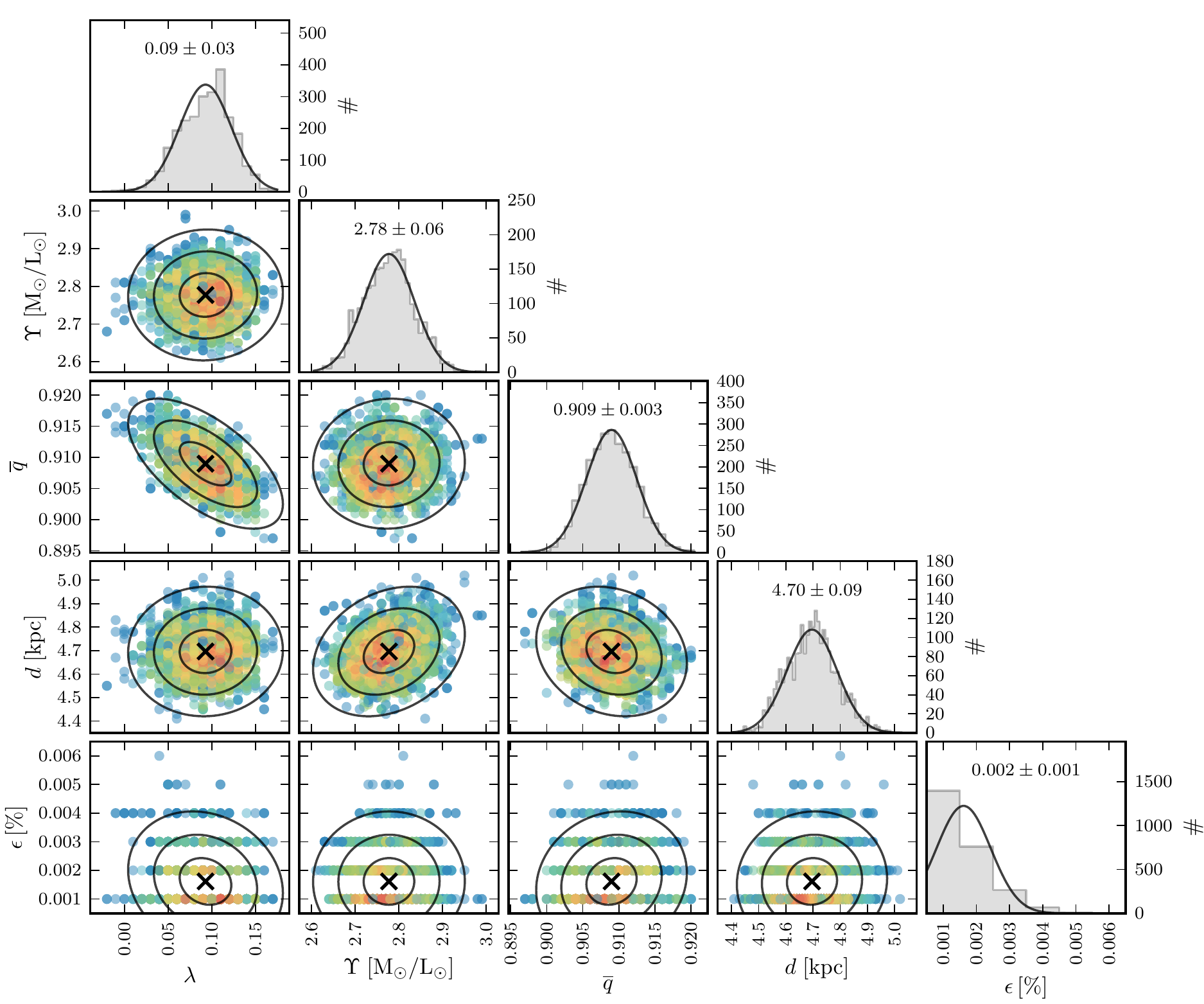}
    \caption{MCMC post-burn distributions for dataset A.  The scatter plots show the projected two-dimensional distributions, with ellipses representing the $1\sigma$, $2\sigma$ and $3\sigma$ regions of the projected covariance matrix.  The histograms show the projected one-dimensional distributions, with lines representing the $1\sigma$ projected covariance matrix.  From top-to-bottom and left-to-right, the panels show anisotropy parameter $\lambda = - \ln \left( 1 - \beta \right)$, mass-to-light ratio $\Upsilon$, median intrinsic flattening $\qbar$, distance $d$ and contamination fraction $\epsilon$.  The anti-correlation between $\lambda$ and $\qbar$ reflects the shape-anisotropy degeneracy.  The discrete nature of the contamination-fraction panels is due to rounding that we employ in our modelling.}
    \label{fig:seta_dbns}
\end{center}
\end{figure*}

We begin by applying our models to a dataset that has been cleaned of possible contaminants - set A (see \autoref{table:datasets}).  \autoref{fig:seta_evol} shows the evolution and eventual convergence of the MCMC chain.  The coloured points show the values sampled by the walkers at each step with the colours representing the likelihood of the model, (red high and blue low).  The solid lines show the means of the walker values and the dotted lines show the 1$\sigma$ dispersions of the walker values.  All of the parameters converge tightly; the contamination fraction converges first, and then the other parameters follow.

Our MCMC chains run for 250 steps.  We consider the first 200 steps as the burn-in phase that finds the region of parameter space where the likelihood is highest.  The final 50 steps then constitute the post-burn phase that explores the high-likelihood region.  \autoref{fig:seta_dbns} shows the post-burn distributions for dataset A.  We do not show the points from all 50 steps, but only those from every second step as MCMC chains have a one-step memory - that is, points from the step $n$ are correlated with the points from the step $n-1$.  The scatter plots show the two-dimensional distributions of the parameters, with points coloured according to their likelihoods (red high and blue low).  The ellipses show the $1\sigma$, $2\sigma$ and $3\sigma$ regions of the covariance matrix for the post-burn parameter distribution, projected into each 2-d plane.  Anisotropy parameter $\lambda$ anti-correlates with the mean flattening $\qbar$ -- this is the shape-anisotropy degeneracy.  The histograms show the one-dimensional distributions of the parameters; the solid black lines show the one-dimensional mean and standard deviation.  The histogram panels also give the one-dimensional mean and uncertainty for each of the parameters.  We deliberately selected the dataset to be contaminant free, and we see from \autoref{fig:seta_evol} that we were successful as the contamination fraction $\epsilon$ has converged to zero with very little scatter.

\begin{table*}
    \caption{The post-burn one-dimensional parameter estimates and uncertainties for all six datasets.  For reference, the values we expect based on the study of \citet{vandeven2006} are shown in the final row with the mass-to-light ratio corrected according to \citet{dsouza2013}.  Dataset A has all cuts applied; dataset B relaxes the LV cuts; dataset C further relaxed the PM membership cuts; datasets D and E relax the PM blending cuts and errors cuts respectively; dataset F has not cuts applied.  For more details of the datasets, see \autoref{sect:data_select} and \autoref{table:datasets}.  We have highlighted the results for dataset C as that is our preferred dataset (see \autoref{sect:bestmodel}).}
    \label{table:results}
    \begin{center}
    \begin{tabular}{cccccccc}
        \hline
        \hline
        dataset & $\lambda$ & $\beta$ & $\Upsilon$ ($\Msun / \Lsun$)
            & $\qbar$ & $i$ (deg) & $d$ (kpc) & $\epsilon (\%)$ \\
        \hline
        expected & $\sim 0$ & $\sim 0$ & $2.7 \pm 0.1$ & $0.91 \pm 0.01 $
            & $50 \pm 4$ & $4.8 \pm 0.3$ & - \\
        \hline
        A & $0.09 \pm 0.03$ & $0.09 \pm 0.03$ & $2.78 \pm 0.06$
            & $0.909 \pm 0.003$ & $50.16^{+1.31}_{-1.20}$
            & $4.70 \pm 0.09$ & $0.002 \pm 0.001$ \\
        B & $0.11 \pm 0.02$ & $0.10 \pm 0.02$ & $2.70 \pm 0.05$
            & $0.908 \pm 0.004$ & $49.87^{+1.29}_{-1.18}$
            & $4.61 \pm 0.08$ & $0.190 \pm 0.012$ \\
        \textbf{C} & $\mathbf{0.11 \pm 0.02}$ & $\mathbf{0.10 \pm 0.02}$
            & $\mathbf{2.71 \pm 0.05}$ & $\mathbf{0.908 \pm 0.004}$
            & $\mathbf{49.76^{+1.28}_{-1.18}}$ & $\mathbf{4.59 \pm 0.08}$
            & $\mathbf{0.259 \pm 0.013}$ \\
        D & $0.12 \pm 0.03$ & $0.11 \pm 0.02$ & $2.84 \pm 0.05$
            & $0.910 \pm 0.003$ & $50.69^{+1.27}_{-1.17}$
            & $4.37 \pm 0.07$ & $0.259 \pm 0.014$ \\
        E & $0.09 \pm 0.02$ & $0.09 \pm 0.02$ & $3.04 \pm 0.05$
            & $0.908 \pm 0.003$ & $49.87^{+1.03}_{-0.96}$
            & $4.09 \pm 0.07$ & $0.404 \pm 0.015$ \\
        F & $0.14 \pm 0.02$ & $0.13 \pm 0.02$ & $3.56 \pm 0.05$
            & $0.917 \pm 0.002$ & $53.14^{+0.99}_{-0.93}$
            & $3.47 \pm 0.05$ & $0.406 \pm 0.015$ \\
        \hline
        \hline
    \end{tabular}
    \end{center}
\end{table*}

The one-dimensional means and uncertainties of the parameters are shown in \autoref{table:results}, along with the results expected based on previous studies that fitted velocity moments after binning.  The estimates we obtain are all in agreement with the values that we expected, so our models appear to work well on the cleaned data.

\subsection{Relaxing the data-selection cuts}

\begin{figure*}
\begin{center}
    \includegraphics[width=\linewidth]{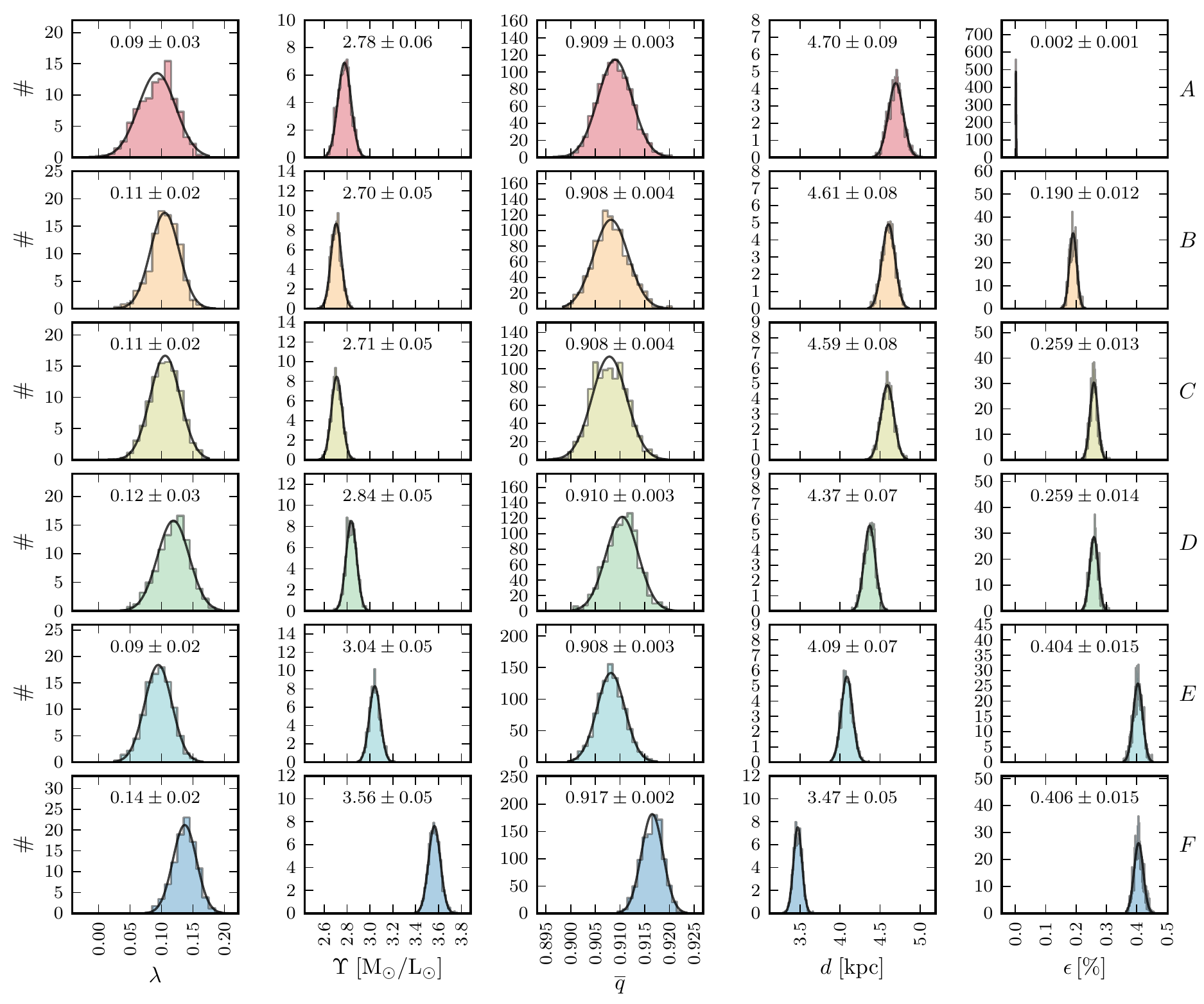}
    \caption{MCMC post-burn distributions for datasets A (top row) through F (bottom row).  The histograms show the projected one-dimensional distributions, with lines representing the $1\sigma$ projected covariance matrix.  From left-to-right, the panels show anisotropy parameter $\lambda = - \ln \left( 1 - \beta \right)$, median intrinsic flattening $\qbar$, mass-to-light ratio $\Upsilon$, distance $d$ and contamination fraction $\epsilon$.  The values of the mean and 1-sigma uncertainties are also given in each panel.  Set A is a cleaned sample and returns parameters consistent with previous studies.  Adding line-of-sight velocity non-members or stars with high errors (set B) and further adding proper motion non-members (set C) recovers the same parameters.  Adding proper motion blended stars (set D), proper motion high-error stars (set E) or both (set F) causes the predicted mass-to-light ratio to increase and the predicted distance to decrease; this indicates that the errors on the blended and high-error proper motion stars have been underestimated.  As expected, the contamination fraction changes for each dataset as relaxing the cuts changes the percentage of contaminants in each sample.}
    \label{fig:all_hists}
\end{center}
\end{figure*}

Now we have established that our framework is successful when applied to a set of bona fide members, it is time to start relaxing the data-selection cuts to include contaminating populations and high-error stars.  \autoref{fig:all_hists} shows the one-dimensional post-burn parameter distributions for all six datasets (from A at the top through to F at the bottom), with the columns showing, from left to right: anisotropy parameter $\lambda$, mass-to-light ratio $\Upsilon$, mean flattening $\qbar$, distance $d$ and contamination fraction $\epsilon$.  The one-dimensional parameter means and uncertainties are also indicated in each panel, as well as collected in \autoref{table:results}.

For all parameters and all datasets the distributions are gaussian, implying that all six runs converged well on their region of preferred parameter space.  We have shown all parameters on the same scale to facilitate comparison of parameter estimates across different datasets.  We can see that datasets B and C estimate parameters in excellent agreement with our clean dataset A, while datasets D, E and F estimate high mass-to-light ratios and correspondingly low distances.  Also from the final column we see that increasing the number of contaminants present in the dataset increases the estimated contamination fraction, as expected.  The overall trends are clear, now let us consider each dataset in turn.

\textbf{Set B}: Contamination fraction $\epsilon$ has increased here, reflecting the fact that relaxing the selection cuts on the line-of-sight velocity sample introduced contaminants to the sample.  However, it is encouraging to see that the other values are all in agreement with those obtained using dataset A, and with the values that we expected from previous studies.  Our models have handled the inclusion of line-of-sight contaminants very well.

\textbf{Set C}: Contamination fraction $\epsilon$ has increased further as relaxing the velocity-selection cuts on the proper-motion sample has included more contaminants.  The other values are all in agreement with those obtained using datasets A and B, and with the values that we expected from previous studies.  Once again, the models have handled the proper-motion foreground population successfully.

\textbf{Set D}: Although the anisotropy and flattening are consistent with the previous datasets and previous studies, the distance we obtain here is lower than we expect and the mass-to-light ratio is much higher so it appears that adding in blended stars from the proper-motion dataset is unsuccessful.  We note that the addition of the blended stars has not increased the contamination fraction; so the blended stars are predicted to be cluster members and not contaminants, however their inclusion has returned a set of incorrect parameter estimates.

\textbf{Set E}: Once again, the distance is lower and the mass-to-light ratio much higher than we expect, although the anisotropy and flattening are consistent with our expectations.  Including proper-motion stars with high errors has been unsuccessful.

\textbf{Set F}: As including blended stars (set D) and high-error stars (set E) separately decreased the distance and increased the mass-to-light ratio favoured by the models, we might expect that adding blended stars and high-error stars together (set F) would drive the distance down even lower and the mass-to-light ratio up even higher, and that is, indeed, what we see here.  The inclination angle has also increased for this dataset.  Again, we note that adding blended stars to set E to obtain set F has not changed the estimate of the contamination fraction, indicating that the blended stars are cluster members.

Adding in blended stars and high-error stars from the proper-motion dataset is clearly not successful.  However, we do not believe this to be a failure of our models, as the models for datasets B and C demonstrated that they are able to work well even when a contaminating population is included.  Furthermore, as we discussed above, all runs are able to converge on a best model.  So the failure of these models is not because the chain is unable to converge, instead it is because the chain converges to incorrect mass-to-light ratios and distances as a direct consequence of the additional data that has been included.  This suggests that the systematic errors on the blended stars and the high-error stars in the proper-motion dataset have been underestimated.

Our models assume that the errors on the proper motion measurements are uncorrelated, which may well not be true, however, in the absence of correlation estimates, this assumption is the best we can do.  We tried to account for the unreliability of the blended stars and high-errors stars by adding a (constant) systematic error in quadrature to their quoted uncertainties. For such an approach to be successful (i.e. to return parameters consistent with those expected), the required systematic error was high, indicating that the model was effectively disregarding these stars.  Similarly, we tried including these unreliable stars with a (constant) lower weight in our analysis.  The weight required was low, once again indicating that the model was effectively disregarding these stars.  This highlights the fact that accurate errors are vital for studies of this nature.

\subsection{Best model}
\label{sect:bestmodel}

\begin{figure*}
\begin{center}
    \includegraphics[width=\linewidth]{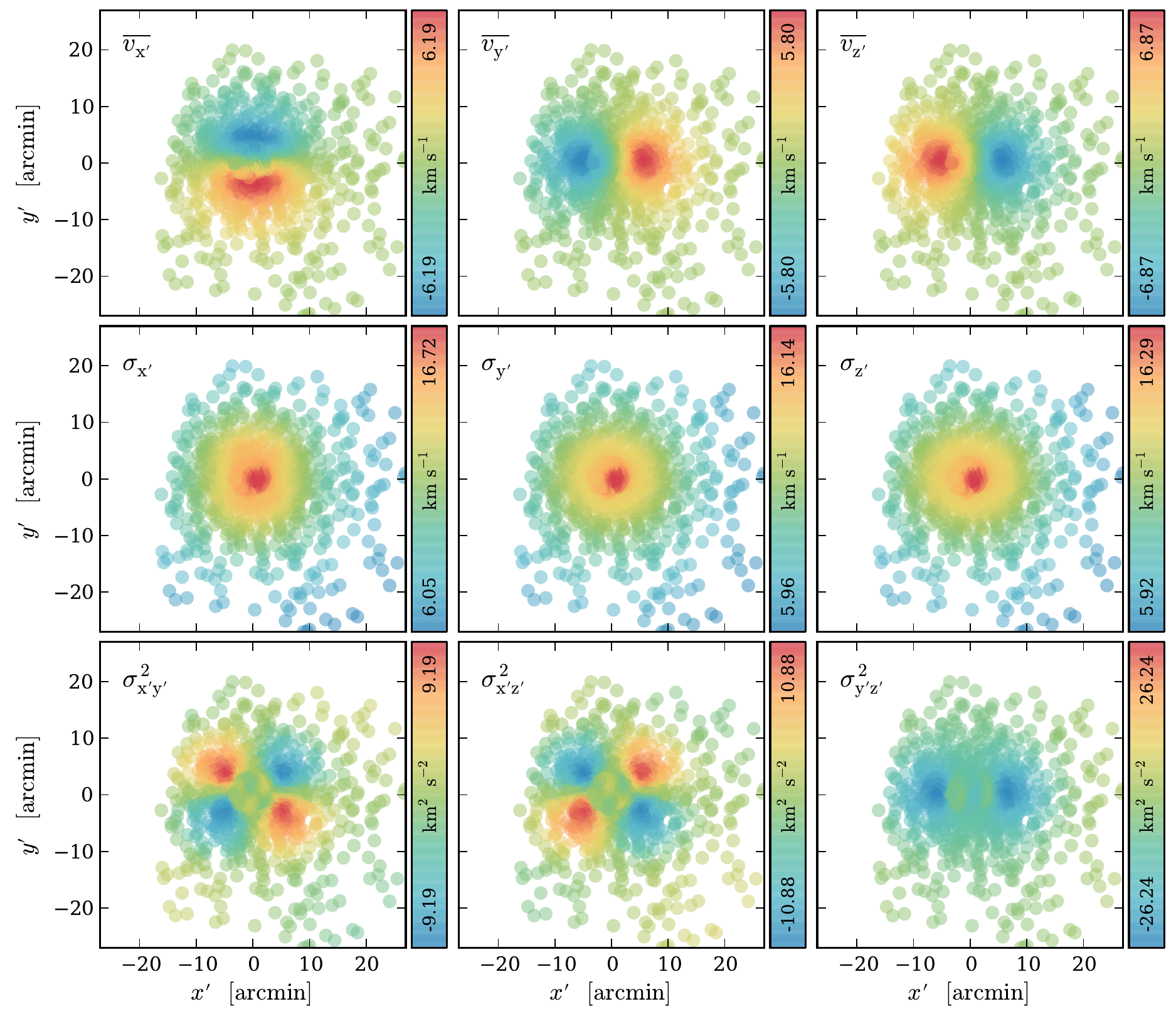}
    \caption{Predicted velocity and dispersion maps for the best model; each point represents the position of a star in our dataset (C) and the colour of the point indicates the value of the velocity or dispersion predicted at that position.  The quantity plotted is shown in the upper-left corner of each panel.  The velocities (top row) show significant rotation and the dispersions (middle row) highlight the very high central velocity dispersion of \wcen{}.  The covariances (bottom row) are clearly non-zero, demonstrating the importance of their inclusion in our likelihood calculations.}
    \label{fig:bestmodel}
\end{center}
\end{figure*}

We believe the velocity errors have been underestimated for the proper motion stars for which there is some blending or for which the errors are large.  As such, we favour the models run using dataset C; recall that to extract this dataset we made no cuts on velocity, only on the degree on blending and on the magnitude of the velocity uncertainties.  In this way we eliminate stars for which we do not trust the quoted errors, while still allowing the models to identify any outliers among the remaining stars.  Here we will concentrate on the dataset C parameter distributions and models.

As we have used the same dataset as \citet{vandeven2006}, we expected to find properties in very good agreement with that study.  Thus, so far, we have focused on comparing our results with the \citet{vandeven2006} study; and simply noted that these results are in good agreement with previous and subsequent studies.  However, it is worth taking a moment here to verify that claim and consider our results in a broader context.

\citet{seitzer1983} estimated a mass-to-light ratio of $2.3 \; \Msun / \Lsun$, while \citet{meylan1987} estimated $2.9 \; \Msun / \Lsun$.  The mass-to-light ratio has been estimated as high as $4.1 \; \Msun / \Lsun$ \citep{meylan1995}, however this is likely an overestimate as this study used spherical models, yet we know that \wcen{} is significantly flattened.  \citet{harris1996} found a distance for \wcen{} of $5.0 \pm 0.2$~kpc, which is the value we adopted for part of our data processing in \autoref{sect:data_process}.  \citet{thompson2001} estimated a distance $5.36 \pm 0.30$~kpc using an eclipsing binary in \wcen{} and \citet{delprincipe2006} found $5.5 \pm 0.1$~kpc using RR Lyrae stars.

More recently, \citet{vandermarel2010} found a mass-to-light ratio of $2.64 \pm 0.03 \; \Msun / \Lsun$ and distance of $4.70 \pm 0.06$~kpc for models with no IMBH and mass-to-light ratio $2.59 \pm 0.03 \; \Msun / \Lsun$ and distance $4.75 \pm 0.06$~kpc for models including an IMBH.  \citet{noyola2008} and \citet{noyola2010} adopted the distance from \citet{vandeven2006} for their models, and found a mass-to-light ratio $\sim 2.7 \; \Msun / \Lsun$ outside of the core radius, with a much higher central mass-to-light ratio $\sim 6.7 \; \Msun / \Lsun$, which they attribute to the presence of an IMBH.

In this study, we have estimated a mass-to-light ratio of $2.71 \pm 0.02 \; \Msun / \Lsun$ and a distance of $4.59 \pm 0.08$~kpc.  We have already noted that these are in good agreement with the \citet{vandeven2006} study and now we can see that they are also consistent with many other previous studies.

\begin{figure*}
\begin{center}
    \includegraphics[width=0.64\linewidth]{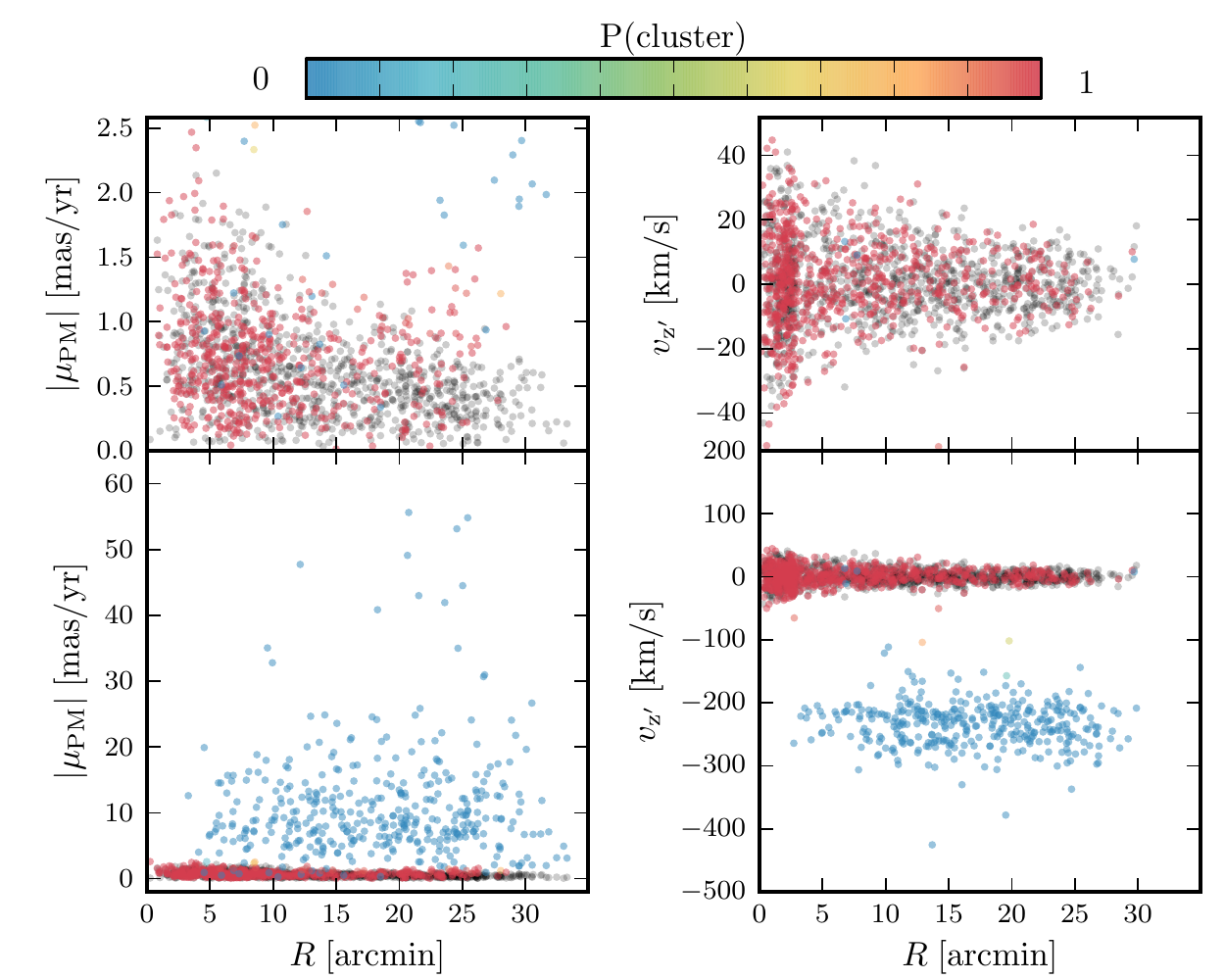}
    \includegraphics[width=0.32\linewidth]{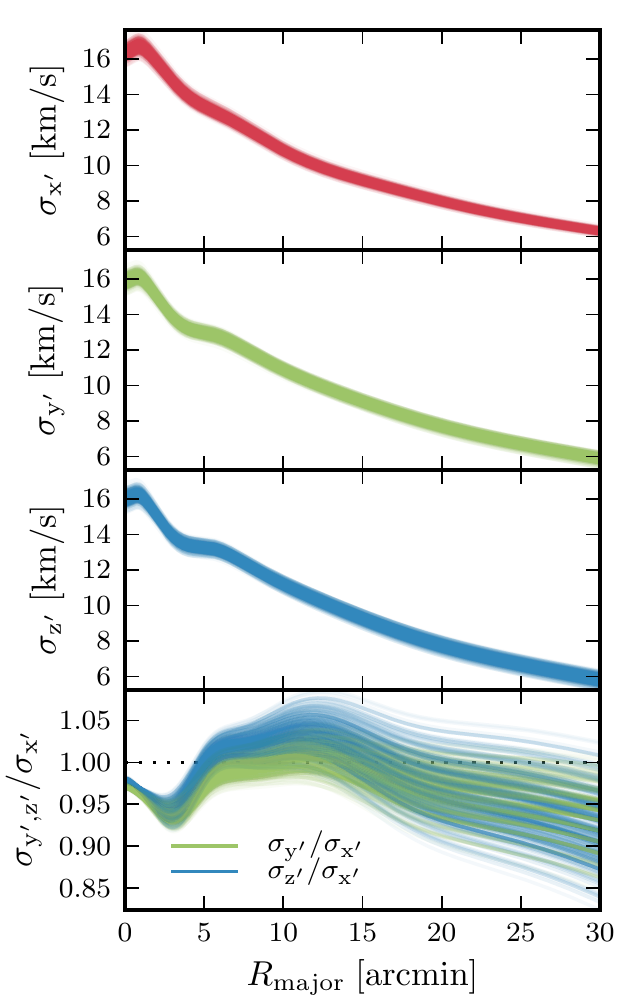}
    \caption{\textit{Left:} A comparison of the observed and predicted proper motions (left) and line-of-sight velocities (right) as a function of projected distance for dataset C.  The lower panels show the full position-velocity plane, and the upper panels zoom in to the velocity range of the model predictions.  In each panel, the black points show velocities drawn from the model distributions at each position and the coloured points show the data.  The colours represent the posterior membership probability $\pmi$ of the stars calculated via \autoref{eqn:postmem}; we can see that most stars have a posterior probability of $\sim1$ (red) or $\sim0$ (blue) with very few intermediate values, indicating that the models are very successful in identifying outliers.  The coherence of the model (black) and high-membership-probability ($\sim$red) points demonstrates the good agreement between data and model.  In order to aid visualisation, we show all stars with $\pmi < 0.99$ or $R > 20$~arcmin, and every fifth star from the remaining population.  \textit{Right:} Velocity dispersion and anisotropy profiles along the major axis for a subset of models from the post-burn MCMC sample.  The top (upper middle, lower middle) panel shows the velocity dispersion in the direction of the major axis (minor axis, line of sight).  The bottom panel shows the minor-axis to major-axis anisotropy in green and the line-of-sight to major-axis anisotropy in blue.  The dotted line indicates isotropy.  There is very little scatter between the different models, suggesting that the estimated `best' parameters are a good representation of the larger model distribution.}
    \label{fig:modeldata}
\end{center}
\end{figure*}

The beauty of MCMC parameter estimation is that we obtain not just one best-fit value (with a one-dimensional uncertainty) for each of our free parameters, but instead a full N-dimensional parameter distribution (where \autoref{fig:seta_dbns} shows a set of one- and two-dimensional projections of that parameter space).  Nevertheless, sometimes it is useful to consider a `best' model as a representative of the wider distribution of models.  We consider our best model to be that with the mean parameters from the final distribution, as given in \autoref{table:results}.  That is, anisotropy parameter $\lambda = 0.11$, mass-to-light ratio $\Upsilon = 2.71 \; \Msun / \Lsun$, median flattening $\qbar = 0.909$, distance $d = 4.59$~kpc and contamination fraction $\epsilon = 0.00259$.

To illustrate the dynamical state of \wcen{}, \autoref{fig:bestmodel} shows predicted velocity and dispersion maps for the best model.  Each point represents the position of a star in our dataset (C) and the colour of the point indicates the value of the velocity or dispersion predicted at that position, as indicated by the colour bars.  The top row shows the major-axis, minor-axis and line-of-sight velocities; all three panels show significant rotation.  The middle row shows the major-axis, minor-axis and line-of-sight velocity dispersions, which highlight the very high central velocity dispersion of \wcen{}.  The bottom panels show the covariances; while reasonably small, they are certainly non-zero and should not be neglected in our likelihood calculations.

Of course, the best model may not necessarily be a good model, and so we now compare the model predictions against the data to show that the model does indeed reproduce the data well.  \autoref{fig:modeldata} shows the observed proper motions (left-hand panel) and observed line-of-sight velocities (middle panel) for dataset C as a function of projected distance.  Our dataset is large so, in order to aid visualisation, we show all stars with $\pmi < 0.99$ or $R > 20$~arcmin, and every fifth star from the remaining population.  The points are coloured according to the posterior membership probability $\pmi$ of the stars calculated via \autoref{eqn:postmem}.  Note that most stars have a posterior probability of $\sim1$ (red) or $\sim0$ (blue) with very few intermediate values, indicating that the models are very successful in identifying outliers.  Now for each star in dataset C, we draw a proper motion and a line-of-sight velocity from the model distribution at the position of the star; we show these model predictions as black points.  As the range of the observed velocities is much larger than the predicted velocities (as we have not excluded velocity outliers in our dataset), we show the full position-velocity plane in the lower panels and then zoom in to the velocity range of the model predictions in the upper panels.  The coherence of the model (black) and high-membership-probability data ($\sim$red) demonstrates the good agreement between data and model.

Thus far we have considered a single `best' model although, as we have discussed, the MCMC process actually gives us a distribution of good models.  So we also wish to consider how this best model compares to other models in the post-burn sample of the MCMC chain.  To that end, the right-hand panels of \autoref{fig:modeldata} show velocity dispersion and anisotropy profiles along the major axis for a subset\footnote{We show only a subset for clarity; the subset was chosen by first selecting only every second run (because MCMC chains have a one-step memory, as previously discussed) and then selecting the final five of these runs for a total of 500 models.} of the post-burn MCMC sample.

The top, upper middle and lower middle panels show the velocity dispersion in the direction of the major axis, minor axis and line of sight respectively.  There is very little scatter among the 500 models represented here, indicating that the best model is representative of the models in the post-burn sample.  The bottom panel shows the minor-axis to major-axis anisotropy in green and the line-of-sight to major-axis anisotropy in blue.  The dotted line indicates isotropy.  There is more scatter here as we are now taking the ratio of two similar numbers, nevertheless the shapes of the anisotropy profiles are in good agreement.  We do not include a data comparison in these plots, as to do so we would have to clean the dataset of contaminants, bin the remaining sample and then calculate dispersions in each bin; this is exactly what we have been trying to avoid by developing these models.

% ---------------------------------------------------------------------------- %

\section{Discussion}
\label{sect:discuss}

In the previous sections, we have shown that a discrete approach to dynamical modelling is able to recover the properties of \wcen{} remarkably well.  It is now worth considering how we might extend the basic models we have presented here.

We have adopted a single velocity anisotropy and mass-to-light ratio for the entire system.  This is a reasonable assumption for the present work as \citet{vandeven2006} and \citet{vandermarel2010} both showed that the change of anisotropy and mass-to-light ratio with radius is mild.  However, if we wish to investigate the possible presence of unseen matter, the models would benefit from greater freedom in the anisotropy and mass profiles.  We have allowed the rotation parameters to vary across the cluster, but we did not include these as free parameters in our models, instead fixing them to the values estimated by \citet{dsouza2013}.  A better approach would then be to allow the anisotropy, rotation and mass-to-light ratio to vary across the MGE, resulting in nonparametric anisotropy, rotation and mass-to-light profiles \citep[see also][]{vandenbosch2006, jardel2012, denbrok2013}.  The one-dimensional gaussian background models we have used here are also overly simplistic; certainly for objects in the Milky Way, the models could be improved by estimating the foreground and background contamination using, for example, the Besan\c{c}on models of \citet{robin2003}.

For globular clusters, like \wcen{}, there is also considerable debate about the presence (or absence) of IMBHs at their centres.  Our models are readily extensible to include IMBHs: we can simply add an extra gaussian to the mass MGE to simulate this non-luminous massive component.  In this way, we hope to determine whether or not globular clusters host IMBHs, and to place limits on the mass of the IMBHs, if they are found to exist.  Such studies require large amounts of data near the projected centre of the cluster.  Of course, it is only stars that are \textit{physically} close to the centre of the cluster that are sensitive to an IMBH.  Discrete modelling is important to ensure that we make optimal use of the data at the projected centre \citep[see][]{denbrok2013}.

Another matter of debate for some globular clusters, including \wcen{}, is whether or not they contain dark matter, which has implications for theories of their formation.  If they are indeed the stripped remnants of nucleated dwarf ellipticals as has been suggested, then they would once have contained dark matter; most of this dark matter will have been stripped away, but small amounts could still be detectable.  Of course, for dwarf spheroidal galaxies, there is no doubt of their dark matter content.  We know that they are some of the most dark-matter-dominated systems that exist, but we do not know how that dark matter is distributed.  Cosmological simulations predict cuspy profiles while observations tend to favour cored profiles, though modellers are typically able to find both cored and cusped profiles that can describe the data.  In light of this, we would also like to add dark matter to our models.  Just as we discussed for the inclusion of IMBHs, we can fit for an extra dark matter component by adding extra gaussians to the MGE that approximate any dark matter profile we may wish to test.

The data we have used in this study is not sufficient to draw conclusions on a possible IMBH or dark matter component in \wcen{}, however there is more data available for \wcen{} than we have used here, which will enable us to do so in subsequent studies.  \citet{anderson2010} presented almost 170\,000 HST proper motions near the projected centre of \wcen{}, with which we hope to search for an IMBH; similar datasets will soon be available for many more clusters \citep{bellini2013}.  In addition, \citet[][]{bellini2009} provided almost 360\,000 ground-based proper motions out to 33~arcmin, with which we hope to study the outermost reaches of \wcen{} for evidence of dark matter.

A further extension that can be made to these models is the inclusion of non-kinematic data, particularly abundance information.  Dwarf spheroidal galaxies and an increasing number of globular clusters are found to host multiple stellar populations.  \wcen{} is a prime example: \citet{johnson2010} found that its red giant branch (RGB) metallicity distribution is best described by five individual, overlapping components.  The origin and formation of these populations is unknown.  We would like to know if stars of different metallicity also have different kinematical properties, which could constrain their formation histories.

The simplest way to do this is to split a sample of stars into a metal-rich and a metal-poor sample and to study their kinematics separately.  This has been done before for binned models of Sculptor \citep[][]{battaglia2008}.  The problem with hard cuts that split the data is that the size of the datasets is reduced; of course, this is more of a problem for binned models where the datasets have already been severely degrading through the binning itself, but may also be a challenge for discrete models, particularly when sample sizes are small to start with.  Hard cuts also require that we fix the boundary between metal-rich and metal-poor stars, which is not straightforward as metallicity distributions are overlapping.  The great power of discrete models is that instead of making hard cuts, we can extend the likelihood functions to incorporate metallicity information and let the models determine which stars are more likely to be metal rich and which are more likely to be metal poor \citep[see also][]{amorisco2012b, amorisco2013}.

This discussion also highlights another issue that we should consider.  \wcen{} is one of the best datasets that we have for globular clusters or dwarf galaxies.  We have proper-motion and line-of-sight-velocity datasets, both numerous and to high precision.  For other Local Group objects, where the data is neither as plentiful or of such high quality, what constraints will we be able to place on their properties?  Conversely, how many (or how few) stars do we need, and of what quality, in order to determine their structures?

Finally, we consider how we might extend the modelling machinery itself.  In this paper, we have used Jeans models to calculate velocity moments for a set of model parameters.  The particular Jeans models we have used are axisymmetric, though we note that this is not a drawback of Jeans models in general as solutions of the Jeans equations in triaxial geometry are available \citep{vandeven2003}.  For \wcen{}, axisymmetry is a reasonable assumption, however triaxial models would be a better choice for dwarf spheroidal galaxies as dark matter halos are predicted to be (prolate) triaxial \citep{jing2002}.  Even so, to solve the Jeans equations in general (ad-hoc) assumptions on the velocity anisotropy have to be made, and, of greater concern, Jeans models can return unphysical solutions with negative distribution functions. Starting from a (parameterised) non-negative distribution function is also very hard; even though its dependence on six phase-space coordinates can in general be reduced to three integrals of motion through Jeans' theorem \citep{jeans1915}, only for specific choices like St\"ackel potentials explicit expression of all integrals of motion are known \citep[e.g.][]{dezeeuw1985}. Numerical modelling techniques such as \citet{schwarzschild1979}'s orbit-superposition \citep[e.g.][]{rix1997, cretton1999, thomas2004, vandeven2006, vandenbosch2008} and made-to-measure \citep[M2M, e.g.][]{syer1996, delorenzi2007, dehnen2009, long2010} methods return a physical solution without having to specify the distribution function and hence also without having to make any (ad-hoc) assumptions on the velocity anisotropy. Since the drawback of these numerical techniques is that they can be computationally expensive, the Jeans models will still be very useful to first reduce the larger possible parameter space.

% ---------------------------------------------------------------------------- %

\section{Conclusions}
\label{sect:conc}

We have developed a discrete dynamical modelling framework and have successfully applied it to Galactic globular cluster \wcen{}.

By treating the stars as discrete data points, we do not suffer from the loss of information inherent in analyses that bin the stars and calculate velocity moments in each bin.  We use Jeans models to calculate the predicted velocity and dispersion for a star under a given set of model parameters.  We allow for a contaminating population in the models, instead of eliminating suspected non-member stars from the datasets a priori.  Finally, we adopt a maximum-likelihood approach to evaluate how well a model is able to reproduce the data and use MCMC to efficiently explore our parameter space.

We tested our models on \wcen{} as it is a scientifically interesting object for which a large quantity of high-quality data is available.  We were able to recover parameters consistent with previous modelling attempts \citep[e.g.][]{vandeven2006, vandermarel2010}, even in the presence of a contaminating population.  We find that \wcen{} has a mildly radial velocity anisotropy $\beta = 0.10 \pm 0.02$, an inclination angle of $i = 50^\circ \pm 1^\circ$, a V-band mass-to-light ratio $\Upsilon = 2.71 \pm 0.05 \; \Msun / \Lsun$ and is at a dynamical distance $d = 4.59 \pm 0.08$~kpc.  Our models have not considered the possibility of an IMBH at its centre or of dark matter in its outer regions, however they are readily extensible to do so, and we plan to revisit these issues in future papers.

In a similar study with the same datasets, \citet{vandeven2006} found that stars from the \wcen{} proper motion catalogue of \citet{vanleeuwen2000} with average errors larger than 0.2~\masyr{} inflated the velocity dispersion (below this error limit, the velocity dispersion remained constant).  As a result, they removed these stars from their analysis.  They also removed stars that were blended in the photographic plates.  We ran models both with and without these low-quality (high-error or blended) stars.  Models that included the low-quality stars returned a lower best-fit distance and a higher best-fit mass-to-light ratio than we obtained for models of the high-quality stars (which were in excellent agreement with previous studies).  The failure of our models when including the low-quality stars is not a fault of the models but of the data and highlights the importance or proper error estimation.

This is a promising start.  However, in this preliminary analysis, we have used models that require a number of undesirable assumptions, we have used only simple background models and we have not included any chemical information.  Nevertheless, these results demonstrate that we have the machinery in place to handle both current and upcoming datasets in the Local Group, now we can work on further developing the maximum likelihood techniques to work with more powerful dynamical models and to incorporate more than only velocity information.  We have shown here that our discrete models can successfully reproduce results obtained from previous binned models.  The true advantages of such a discrete treatment of these datasets will become apparent in future papers as we extend our models beyond that which is possible with binned data.

% ---------------------------------------------------------------------------- %

\section*{Acknowledgements}

LLW wishes to thank Nicolas Martin, David Hogg and Coryn Bailer-Jones for interesting discussions on the matter of maximum likelihood methods, Tim de Zeeuw for feedback on the draft, and Tom Robitaille for advice regarding code publishing.  We also thank the referee Eva Noyola for the very helpful report that improved the presentation of the paper.  This work was supported by Sonderforschungsbereich SFB 881 ``The Milky Way System" of the German Research Foundation (DFG).

% -----------------------------------------------------------------------------

% set bibliography directory
\bibliographystyle{mn2e}
\bibliography{refs}

% -----------------------------------------------------------------------------

\appendix

\section{JAM Calculations}
\label{sect:jamapp}

Here we present a complete derivation of all the first and second velocity-moment calculations in the JAM formalism.  \citet{cappellari2008} originally calculated the line-of-sight first and second moment equations, \citet{dsouza2013} calculated the second moments for the major- and minor-axis proper motions, and \citet{cappellari2012} calculated the second moment cross-terms.  However, the first moments for the major- and minor-axis proper motions remain uncalculated, so we do so here.  For completeness, we include a derivation of all the first- and second-moment equations.  We provide only a brief introduction of the models and concentrate on the calculations; for an extended discussion on the JAM formalism, see \citet{cappellari2008}.  Our code, written in C, is available at http://github.com/lauralwatkins/cjam.

\subsection{Coordinate system}

As we are using axisymmetric models of dynamical systems, it is natural to use either cartesian $(x, y, z)$ or cylindrical polar $(R, \phi, z)$ coordinates to describe the intrinsic shape of the system, where the $z$-axis is the symmetry axis and $R^2 = x^2 + y^2$.  The velocity components are then related via
\begin{equation}
    \left( \begin{array}{c}
        \vx \\
        \vy \\
        \vz
    \end{array} \right) = \left( \begin{array}{ccc}
        \cos \phi & - \sin \phi & 0 \\
        \sin \phi & \cos \phi & 0 \\
        0 & 0 & 1
    \end{array} \right) \left( \begin{array}{c}
        \vR \\
        \vphi \\
        \vz
    \end{array} \right) .
\end{equation}

We define a second set of coordinates in the plane of the sky $(x', y', z')$, where the $x'$-axis is aligned with the projected major axis, the $y'$-axis with the projected-minor axis and the $z'$-axis lies along the line-of-sight such that the line-of-sight vector is positive in the direction away from us.  Note that, while the intrinsic coordinates describe a right-handed system, the sky coordinates describe a \textit{left-handed} system because of the way we have defined the positive $z'$-direction.  The sky coordinates related to the intrinsic coordinates via
\begin{equation}
    \left( \begin{array}{c}
        x' \\
        y' \\
        z'
    \end{array} \right) = \left( \begin{array}{ccc}
        1 & 0 & 0 \\
        0 & - \cos i & \sin i \\
        0 & \sin i & \cos i
    \end{array} \right) \left( \begin{array}{c}
        x \\
        y \\
        z
    \end{array} \right) .
\end{equation}
The velocities are related similarly.  Thus, the first velocity moments in the plane of the sky are
\begin{align}
    \vxpbar & = \vRbar \cos \phi - \vphibar \sin \phi \label{eqn:vx} \\
    \vypbar & = - \left( \vRbar \sin \phi + \vphibar \cos \phi \right) \cos i
        + \vzbar \sin i \label{eqn:vy} \\
    \vzpbar & = \left( \vRbar \sin \phi + \vphibar \cos \phi \right) \sin i
        + \vzbar \cos i \label{eqn:vz}
\end{align}
and the second velocity moments are
\begin{align}
    \vxpsqbar = {} & \vRsqbar \cos^2 \phi + \vphisqbar \sin^2 \phi
        \label{eqn:v2x} \\
    \vypsqbar = {} & \left( \vRsqbar \sin^2 \phi + \vphisqbar \cos^2 \phi
        \right) \cos^2 i + \vzsqbar \sin^2 i \nonumber \\
    & - 2 \vRvzbar \sin \phi \sin i \cos i \label{eqn:v2y} \\
    \vzpsqbar = {} & \left( \vRsqbar \sin^2 \phi + \vphisqbar \cos^2 \phi
        \right) \sin^2 i + \vzsqbar \cos^2 i \nonumber \\
    & + 2 \vRvzbar \sin \phi \sin i \cos i \label{eqn:v2z} \\
    \vxpvypbar = {} & \left( - \vRsqbar + \vphisqbar \right) \cos \phi \sin \phi
        \cos i + \vRvzbar \cos \phi \sin i \label{eqn:vxvy} \\
    \vxpvzpbar = {} & \left( \vRsqbar - \vphisqbar \right) \cos \phi \sin \phi
        \sin i + \vRvzbar \cos \phi \cos i \label{eqn:vxvz} \\
    \vypvzpbar = {} & \left( - \vRsqbar \sin^2 \phi - \vphisqbar \cos^2 \phi +
        \vzsqbar \right) \cos i \sin i \nonumber \\
    & - \vRvzbar \sin \phi \left( \cos^2 i - \sin^2 i \right) \label{eqn:vyvz}
\end{align}
where we have included the fact that $\vRbar \vphibar = \vphibar \vzbar = 0$ due to the assumption of axisymmetry.  We will later assume that the velocity ellipsoid is aligned with the cylindrical coordinate system, thus fixing $\vRvzbar = 0$.

\subsection{MGE prescription}
\label{sect:mges}

The JAM models parameterise light and mass profiles as MGEs \citep{emsellem1994}.  In this case, the projected surface brightness $I$ of the object is given by
\begin{equation}
    I \left( x', y' \right) = \sumk \frac{L_k}{2 \pi \sksq q'_k} \exp
        \left[ - \frac{1}{2 \sksq} \left( x'^2 + \frac{y'^2}{q_k^{'2}} \right)
        \right] \label{eqn:mgesurf}
\end{equation}
where, for each of the $N$ gaussian components, $L_k$ is the total luminosity, $0 \le q'_k \le 1$ is the observed (projected) axial ratio and $\sigma_k$ is the dispersion along the major axis.  The corresponding intrinsic (deprojected) luminous density is then
\begin{equation}
    \nu \left( R, z \right) = \sumk \frac{L_k}{ \left( 2 \pi \sksq
        \right)^{\sfrac{3}{2}} q_k} \exp \left[ - \frac{1}{2 \sksq}
        \left( R^2 + \frac{z^2}{q_k^2} \right) \right]
    \label{eqn:nu}
\end{equation}
where the intrinsic axial ratios $q_k$ are related to the projected axis ratios via the inclination angle of the system:
\begin{equation}
    q_k = \frac{\sqrt{q_k^{'2} - \cos^2 i}}{\sin i}
\end{equation}
For face-on systems, $i = 0^{\circ}$; for edge-on systems, $i = 90^{\circ}$.

The mass density of the system is similarly described using a series of $M$ gaussians by
\begin{equation}
    \rho \left( R, z \right) = \sumj \frac{M_j}{ \left( 2 \pi \sjsq
        \right)^{\sfrac{3}{2}} q_j} \exp \left[ - \frac{1}{2 \sjsq} \left( R^2 + \frac{z^2}{q_j^2} \right) \right] .
\end{equation}
In general, the mass gaussians are independent of the luminous gaussians.  This density generates a gravitational potential of
\begin{equation}
    \Phi \left( R, z \right) = - \sqrt{ \frac{2}{\pi} } G \int_0^1 \sumk
        \frac{M_j \hju}{\sigma_j} \dd u
    \label{eqn:phi}
\end{equation}
where $G$ is the gravitational constant and
\begin{equation}
    \hju = \frac{ \exp \left[ - \frac{u^2}{2 \sjsq} \left( R^2 +
        \frac{z^2}{\qju} \right) \right] }{\sqrt{\qju}}
\end{equation}
Extra features, such as black holes and dark halos, can be included in the models by adding extra gaussians to the luminous and mass expansions, as appropriate.

\subsection{Jeans equations}

For an axisymmetric ($\frac{\partial}{\partial \phi} = 0$) system in a steady state ($\frac{\partial}{\partial t} = 0$), the second moment Jeans equations in cylindrical polars are
\begin{align}
    \frac{\nu (\vRsqbar - \vphisqbar)}{R}
        + \frac{\partial (\nu \vRsqbar)}{\partial R}
        + \frac{\partial (\nu \vRvzbar)}{\partial z}
        & = - \nu \frac{\partial \Phi}{\partial R} \\
    \frac{\nu \vRvzbar}{R}
        + \frac{\partial (\nu \vRvzbar)}{\partial R}
        + \frac{\partial (\nu \vzsqbar)}{\partial z}
        & = - \nu \frac{\partial \Phi}{\partial z}
\end{align}
In order to obtain a unique solution for the second moments from these equations, we make two assumptions: that the velocity ellipsoid is aligned with cylindrical polar coordinate system, so that $\vRvzbar = 0$; and that the anisotropy is constant and quantified by $\vRsqbar = b \vzsqbar$.  (Note: when this is the semi-isotropic case when $b = 1$.)  If we further impose the boundary condition that $\nu \vzsqbar = 0$ as $z \to \infty$ then the equations become
\begin{align}
    \nu \vphisqbar (R,z) & = b \left[ R \frac{\partial (\nu \vxsqbar)}
        {\partial R} + \nu \vzsqbar \right] + R \nu \frac{\partial \Phi}
        {\partial R} \\
    \nu \vzsqbar (R,z) & = \int^{\infty}_z \nu \frac{\partial \Phi}
        {\partial z} \dd z .
\end{align}
Substituting for $\nu$ from \autoref{eqn:nu} and $\Phi$ from \autoref{eqn:phi}, we obtain
\begin{align}
    \nuvrsqk = {} & b_k \nuvzsqk \\
    \nuvzsqk = {} & 4 \pi G \int_0^1 \sumj \; \frac{\sksq q_k^2 \nuk q_j
        \rhooj \hju u^2}{\cusq} \df u \\
    \nuvphisqk = {} & 4 \pi G \int_0^1 \sumj \; \left( \dd R^2 + b_k \sksq q_k^2
        \right) \nonumber \\
    {} & \times \frac{\nuk q_j \rhooj \hju u^2}{\cusq} \df u
\end{align}
where $\nuk = \nuk \left( R, z \right)$ and $\rhooj = \rho_j \left( 0, 0 \right)$ and we have defined
\begin{align}
    \cc & = 1 - q_j^2 - \frac{\sksq q_k^2}{\sjsq} \\
    \dd & = 1 - b_k q_k^2 - \left[ (1 - b_k) \cc + (1 - q_j^2) b_k \right] u^2
\end{align}

\subsection{Line-of-sight integration of the second velocity moments}

In general, the second velocity moments have contributions from each of the radial, azimuthal and vertical mean squared velocities, so let us consider
\begin{align}
    I \, \overline{v_{\alpha} v_{\beta}} \left( x',y' \right) = {} & \sumk
        \left[ \intinf \left\{ f_{\alpha \beta} \nuvrsqk + g_{\alpha \beta}
        \nuvphisqk \right. \right. \nonumber \\
    {} & \left. \left. + h_{\alpha \beta} \nuvzsqk \right\} \df z' \right]
    \label{eqn:v2ab}
\end{align}
where $\alpha$ and $\beta$ represent each of the $x'$, $y'$ and $z'$ directions and $f_{\alpha \beta}$, $g_{\alpha \beta}$ and $h_{\alpha \beta}$ are all functions of $z'$.  Substituting from the Jeans' equations, we get
\begin{align}
    I \, \overline{v_{\alpha} v_{\beta}} \left( x',y' \right) = {} & 4 \pi G
        \int_0^1 \left( \sumj \sumk \; \frac{q_j \rhooj u^2}{\cusq} \right. 
        \nonumber \\
    {} & \times \left. \intinf \fabk \; \nuk \hju \df z' \right) \df u
\end{align}
where
\begin{equation}
    \fabk = \left[ \left( f_{\alpha \beta} + g_{\alpha \beta} \right) b_k
        + h_{\alpha \beta} \right] \sksq q_k^2 + g_{\alpha \beta} \dd R^2 ,
    \label{eqn:fabk}
\end{equation}
which is function of $z'$.

Further substituting for the MGE components, we get
\begin{align}
    I \, \overline{v_{\alpha} v_{\beta}} \left( x',y' \right) = {} & \fpigu
        \nqpui \nonumber \\
    {} & \times \vvdenom \nonumber \\
    {} & \times \expabxy \df u
\end{align}
where $\nuok = \nuk(0,0)$ and we have defined
\begin{equation}
    \iabk = \frac{1}{\sqrt{\pi}} \intinf \fabk \exp \left[ -w^2 \right] \df w
\end{equation}
with
\begin{equation}
    w = \rte z' + \frac{\bb \cos i \sin i y'}{\rte}
\end{equation}
and
\begin{align}
    \aa & = \frac{1}{2} \left( \frac{1}{\sksq} + \frac{u^2}{\sjsq} \right) \\
    \bb & = \frac{1}{2} \left( \frac{1 - q_k^2}{\sksq q_k^2} + \frac{(1 -
        q_j^2) u^4}{\sjsq [\qju]} \right) \\
    \ee & = \aa + \bb \cos^2 i .
\end{align}
If $\fabk = \kk_0 + \kk_1 z' + \kk_2 z'^2$ for constants $\kk_0$, $\kk_1$ and $\kk_2$, then
\begin{align}
    \iabk = {} & \kk_0 - \frac{\bb y' \cos i \sin i}{\ee} \kk_1 \nonumber \\
        {} & + \frac{\ee + 2 \bb^2 \cos^2 i \sin^2 i y'^2}{2 \ee^2} \kk_2
\end{align}

\subsubsection{Projected-major-axis proper motion}

Now we consider the particular case for the projected-major-axis proper motion projected second moment along the line-of-sight.  From \autoref{eqn:v2x}, we find
\begin{align}
    f_{x' x'} =  \cos^2 \phi \nonumber \\
    g_{x' x'} = \sin^2 \phi \nonumber \\
    h_{x' x'} = 0
\end{align}
so
\begin{align}
    \ff_{x' x', k} = {} & \dd \sin^2 i \, z'^2 - 2 \dd y' \cos i \sin i \, z'
        \nonumber \\
    {} & + \dd y'^2 \cos^2 i + b_k \sksq q_k^2
\end{align}
and finally
\begin{equation}
    \ii_{x' x', k} = b_k \sksq q_k^2 + \frac{\dd}{\ee^2} \ab^2 y'^2 \cos^2 i
        + \frac{\dd}{2 \ee} \sin^2 i .
\end{equation}

\subsubsection{Projected-minor-axis proper motion}

Now we consider the particular case for the projected-minor-axis proper motion projected second moment along the line-of-sight.  From \autoref{eqn:v2y}, we find
\begin{align}
    f_{y' y'} &= \sin^2 \phi \cos^2 i \nonumber \\
    g_{y' y'} & = \cos^2 \phi \cos^2 i \nonumber \\
    h_{y' y'} & = \sin^2 i
\end{align}
so
\begin{equation}
    \ff_{y' y', k} = \left( b_k \cos^2 i + \sin^2 i \right) \sksq q_k^2
        + \cos^2 i \dd x'^2
\end{equation}
and finally
\begin{equation}
    \ii_{y' y', k} = \left( b_k \cos^2 i + \sin^2 i \right) \sksq q_k^2
        + \dd x'^2 \cos^2 i .
\end{equation}

\subsubsection{Line-of-sight velocity}

Now we consider the particular case for the line-of-sight velocity projected second moment along the line-of-sight.  From \autoref{eqn:v2z}, we find
\begin{align}
    f_{z' z'} & = \sin^2 \phi \sin^2 i \nonumber \\
    g_{z' z'} & = \cos^2 \phi \sin^2 i \nonumber \\
    h_{z' z'} & = \cos^2 i
\end{align}
so
\begin{equation}
    \ff_{z' z', k} = \left( b_k \sin^2 i + \cos^2 i \right) \sksq q_k^2
        + \sin^2 i \dd x'^2
\end{equation}
and finally
\begin{equation}
    \ii_{z' z', k} = \left( b_k \sin^2 i + \cos^2 i \right) \sksq q_k^2
        + \sin^2 i \dd x'^2 .
\end{equation}

\subsubsection{Projected-major-axis proper motion \& projected-minor-axis proper motion}

Now we consider the particular case for the projected-major-axis proper motion and projected-minor-axis proper motion projected second moment along the line-of-sight.  From \autoref{eqn:vxvy}, we find
\begin{align}
    f_{x' y'} & = - \cos \phi \sin \phi \cos i \nonumber \\
    g_{x' y'} & = \cos \phi \sin \phi \cos i \nonumber \\
    h_{x' y'} & = 0
\end{align}
so
\begin{equation}
    \ff_{x' y', k} = - \dd x' y' \cos^2 i + \dd x' \cos i \sin i z'
\end{equation}
and finally
\begin{equation}
    \ii_{x' y', k} = - \frac{\dd}{\ee} x' y' \cos^2 i \ab
\end{equation}

\subsubsection{Projected-major-axis proper motion \& line-of-sight velocity}

Now we consider the particular case for the projected-major-axis proper motion and line-of-sight velocity projected second moment along the line-of-sight.  From \autoref{eqn:vxvz}, we find
\begin{align}
    f_{x' z'} & = \cos \phi \sin \phi \sin i \nonumber \\
    g_{x' z'} & = - \cos \phi \sin \phi \sin i \nonumber \\
    h_{x' z'} & = 0
\end{align}
so
\begin{equation}
    \ff_{x' z', k} = \dd x' y' \cos i \sin i - \dd x' \sin^2 i z'
\end{equation}
and finally
\begin{equation}
    \ii_{x' z', k} = \frac{\dd}{\ee} x' y' \cos i \sin i \ab
\end{equation}

\subsubsection{Projected-minor-axis proper motion \& line-of-sight velocity}

Now we consider the particular case for the projected-minor-axis proper motion and line-of-sight velocity projected second moment along the line-of-sight.  From \autoref{eqn:vyvz}, we find
\begin{align}
    f_{y' z'} & = - \sin^2 \phi \cos i \sin i \nonumber \\
    g_{y' z'} & = - \cos^2 \phi \cos i \sin i \nonumber \\
    h_{y' z'} & = \cos i \sin i
\end{align}
so
\begin{equation}
    \ff_{y' z', k} = \cos i \sin i \left[ (1 - b_k) \sksq q_k^2 - \dd x'^2
        \right]
\end{equation}
and finally
\begin{equation}
    \ii_{y' z', k} = \left[ (1 - b_k) \sksq q_k^2 - \dd x'^2 \right]
        \cos i \sin i
\end{equation}

\subsection{Line-of-sight integration of the first velocity moments}

In general, the first velocity moments will have contributions from each of the radial, azimuthal and vertical velocities.  However, from the assumption that the velocity ellipsoid is aligned with the cylindrical coordinate system, we have $\vRbar = \vzbar = 0$, so let us consider
\begin{equation}
    I \, \overline{v_{\tau}} \left( x',y' \right) = \intinf \nu \, \vphibar
        \; f_{\tau} \df z'.
    \label{eqn:xifirst}
\end{equation}
where $f_{\tau}$ is a function of $z'$ and $\tau$ represents the projected coordinate axes $x'$, $y'$ and $z'$.

For calculation of the second moments, we were forced to make two assumptions in order obtain a unique solution; these are no longer sufficient for a unique solution for the first moments and we require a further assumption.  Physically, we determine (or set) the relative contributions of random and ordered motion to the RMS velocities.  In practice, we are setting a relation between $\vphisqbar$, which we know from the Jeans equations, and $\vphibar$, which we require to calculate the first velocity moment.  To do this, we define for each Gaussian component
\begin{equation}
    \nuvphik = \kappak \left( \nuvphisqk - \nuvrsqk \right)^{\frac{1}{2}}
\end{equation}
where $\kappak$ quantifies the rotation of the $k$th Gaussian component such that $\kappak = 0$ when not rotating and $\left| \kappak \right| = 1$ when the velocity ellipsoid is circular.  Then from
\begin{equation}
    \nu \vphibar^2 = \sumk \left[ \nu \vphibar^2 \right]_k
\end{equation}
we get
\begin{equation}
    \nu \vphibar = \left[ \nu \sumk \kappaksq \left( \nuvphisqk - \nuvrsqk
        \right) \right]^{\frac{1}{2}}
\end{equation}
Substituting this into \autoref{eqn:xifirst}, we obtain
\begin{equation}
    I \, \overline{v_{\tau}} \left( x',y' \right) = 2 \sqrt{\pi G} \intinf
        \ff_{\tau} \gg \df z'
\end{equation}
where $\ff_{\tau} = R f_{\tau}$ and
\begin{equation}
    \gg = \left[ \nu \int_0^1 \sumk \sumj \frac{\kappaksq \nuk
        q_j \rhooj \hju u^2 \dd}{\cusq} \df u \right]^{\frac{1}{2}}.
\end{equation}

\subsubsection{Projected-major-axis proper motion}
Now we consider the particular case for the projected-major-axis proper motion projected first moment along the line-of-sight.  From \autoref{eqn:vx}, we find $f_{x'} = - \sin \phi$ and, thus,
\begin{equation}
    \ff_{x'} = y' \cos i - z' \sin i .
\end{equation}

\subsubsection{Projected-minor-axis proper motion}

Now we consider the particular case for the projected-minor-axis proper motion projected first moment along the line-of-sight.  From \autoref{eqn:vy}, we find $f_{y'} = - \cos \phi \cos i$ and, thus,
\begin{equation}
    \ff_{y'} = - x' \cos i .
\end{equation}

\subsubsection{Line-of-sight velocity}

Now we consider the particular case for the line-of-sight velocity projected first moment along the line-of-sight.  From \autoref{eqn:vz}, we find $f_{z'} = \cos \phi \sin i$ and, thus,
\begin{equation}
    \ff_{z'} = x' \sin i .
\end{equation}

% -----------------------------------------------------------------------------

\label{lastpage}

\end{document}